\documentclass[a4paper,12pt]{article}

\usepackage[utf8]{inputenc}
\usepackage{cancel}
\usepackage{tikz}
\usepackage{ulem}
\usepackage{amsfonts}
\usepackage{amssymb}
\usepackage{graphicx}
\usepackage{amsmath}
\usepackage{enumerate}
\usepackage{mathtools}
\usepackage{subfig}
\usepackage{color}
\usepackage{tikz}
\usepackage{float}
\usepackage{here}
\usepackage{cite}
\usepackage{mathrsfs}
\usepackage{float,epsfig}
\usepackage{dcolumn}
\usepackage{graphicx}
\usepackage{bm}
\usepackage{changepage}
\usepackage{amsmath,amssymb,amsthm}
\usepackage[colorlinks=true,linkcolor=blue,citecolor=red]{hyperref}
\usepackage{booktabs} 
\usepackage{siunitx}  
\tikzset{every picture/.style={line width=0.75pt}} 
\usetikzlibrary{arrows.meta, positioning}
\usepackage{multirow}
\usepackage[toc,page]{appendix}
\usetikzlibrary{arrows.meta}
\usetikzlibrary{bending}
\usetikzlibrary{calc}
\newcommand{\be}{\begin{equation}}
\newcommand{\ee}{\end{equation}}
\newcommand{\bea}{\setlength\arraycolsep{2pt} \begin{eqnarray}}
\newcommand{\eea}{\end{eqnarray}}

\def\0{{\sst{(0)}}}
\def\1{{\sst{(1)}}}
\def\2{{\sst{(2)}}}
\def\3{{\sst{(3)}}}
\def\4{{\sst{(4)}}}
\def\5{{\sst{(5)}}}
\def\6{{\sst{(6)}}}
\def\7{{\sst{(7)}}}
\def\8{{\sst{(8)}}}
\def\sst#1{{\scriptscriptstyle #1}}

\makeatletter \@addtoreset{equation}{section}

\definecolor{lime}{HTML}{A6CE39}

\begin{document}

\title{\Large \textbf{   Constraining  Black Hole Parameters in  Non-Commutative Geometry using Machine Learning }}

\author{\small
Maryem Jemri\thanks{maryem.jemri@um5r.ac.ma. } \\[2mm]
{\small ESMaR, Faculty of Science, Mohammed V University in Rabat, Rabat, Morocco} \\
}

\maketitle

\begin{abstract}
Motivated by string theory, we constrain non-commutative black hole parameters through   shadow  behaviors  using machine learning techniques combined by CUDA computations.  To do so,  we first   investigate the structure of the event horizon of  non-commutative black holes  in the presence of string clouds and dark energy sectors by exploiting CUDA-based methods. We   numerically approach the shadow  properties  and the energy emission rate of   rotating and charged  black holes  in non-commutative geometry via  such high-performance parallel computings.  To bridge these  findings  with observational data, we implement a CUDA-based framework  in order to constrain the involved black  hole  parameters including the non-commutative one.  Using the resulting numerical data, we build  a robust training datasets for a fully connected neural network   to determine whether a given set of parameters matches  with the observational data provided by   Event Horizon Telescope collaborations.  As a result,  we  find that the non-commutative  model under study is consistent with the observations of $SgrA^*_{\mathrm{Keck}}$ black holes. {\noindent}\\
\textbf{Keywords}:   Rotating and charged black holes in non-commutative geometry, EHT collaboration, CUDA numerical codes, Machine learning, fully connected neural network.

\end{abstract}

%

\newpage
\tableofcontents
\newpage
\section{Introduction}
Black hole thermodynamics  including  entropy, temperature, phase transitions, and critical phenomena has received particular importance \cite{ii4,i1,i2}. These studies provide  deep connections between several topics such as  gravity, quantum mechanics, and statistical physics.  Such connections have  played  a leading role in the current effort to develop a unified description of fundamental interactions.

Beyond standard general relativity, black hole solutions have been generalized within the framework of extended theories of gravity and modified geometric structures \cite{i3,i3333}. In these approaches,  the spacetime exhibits additional properties going  beyond the classical Riemannian description.  For instance, such  deformations result from non-commutative (NC) geometry, modified derived structures, and gauge-theory descriptions of gravity, providing new classes of black hole solutions. It has been shown that  the associated  activities  have refined  certain  classical results \cite{ii5}. In particular, the  implementation of NC geometry  in the  spacetime naturally induces modifications to  the Einstein’s field equations \cite{i4,i5,i6}. In addition, extensions such as nonlinear Einstein electrodynamics modify the coupling between geometry and matter fields, giving rise to regular or deformed black hole configurations \cite{i7}. More generally, several effective gravitational theories can also incorporate contributions from dark matter or dark sector effects, which can modify the geometry and  the dynamics of black holes \cite{i8}.  In fact, these approaches provide a  framework for understanding how  spacetime geometric  deformations  and additional physical contributions can  alert  the black hole  physical behaviors.

These theoretical developments have crucial  consequences for observable signatures of black holes. Recently, considerable attention has been devoted to the optical properties of both rotating and non-rotating black holes in deformed spacetimes derived  from various extensions of general relativity \cite{i9,i10,10i,11i,12i,hajar,adil,carlos,belhaj}.  Many models have been investigated including  black hole in Type II superstrings and M-theory where brane objects  have been exploited in the  examination of the associated physical behaviors  \cite{ST1,ST2,ST3,Q22}. It has been revealed that certain  geometric corrections can modify photon trajectories, gravitational lensing effects, and shadow-related observables. Such effects are particularly relevant in empirical activities  reported by the Event Horizon Telescope (EHT) collaboration, which  provided the first direct images of supermassive black hole shadows \cite{ii11,ii12,ii13}. To match  theoretical predictions with observational data, high-performance numerical methods  have been needed \cite{iii11,iii12}. In this context, CUDA has become a powerful platform for parallel computing based on NVIDIA GPUs \cite{o1,o2}. It has been shown that   such numerical computations  enable fast and scalable simulations, significantly reducing computation time while improving numerical accuracy. In black hole physics, these advantages are particularly important for shadow calculations and large parameter spaces investigations  \cite{o3,o4}. Therefore, CUDA-based methods provide an efficient framework for confronting theoretical models with the observational results of the EHT collaboration.

In addition to numerical approaches, certain  data-driven techniques have recently   introduced as powerful tools in black hole physics.  Precisely, machine learning methods provide new ways to extract physical parameters, classify black hole configurations, and improve the reconstruction of observational images. Neural networks, for instance, can estimate accretion parameters from event-horizon-scale images \cite{i12}. Moreover, they could  improve the quality of black-hole images obtained by the EHT international collaboration \cite{i13}. They have also been successfully applied to classify black holes in the mass–spin parameter space \cite{i14}. Furthermore, deep learning techniques have been extended to the analysis of gravitational-wave data, including the estimation of parameters for supermassive binary black holes in simulated LISA observations. These developments supply  the growing role of machine learning as a complementary tool for probing the fundamental properties of black holes and analyzing increasingly complex astrophysical data.

The aim of this work is to  contribute to such activities by constraining   rotating  and charged black hole parameters in NC geometry  via  shadow  properties using machine learning techniques combined with CUDA methods.  Before  matching the obtained  current theoretical predictions with the shadow observations reported by the EHT collaboration, we first discuss the horizon structure of NC black holes  with  string clouds  in the presence of   dark  energy sector by  providing   numerical values using CUDA-based methods.  Then,  we  investigate the shadows and the energy emission rate of  NC rotating and charged black holes with the help of  high-performance parallel computations. To bridge such findings with  observational data, we develop a CUDA-based framework to establish  relevant  constraints on the NC parameter $b$, the cloud of strings parameter $\alpha$, and the quintessence parameter $N$. Finally, we use these numerical results to construct reliable training data for a fully connected neural network (FCNN), in order to  to determine whether a given combination of parameters is consistent with EHT observational constraints. Precisely, we show that the trained FCNN accurately identifies the parameter space regions  corresponding to the $M87^*$, Sgr$A^*_{\mathrm{VLTI}}$, and Sgr$A^*_{\mathrm{Keck}}$ black holes, with predictions evaluated within the $1-\sigma$ and $2-\sigma$ confidence intervals from the EHT collaboration.  As a result,  we find that   the proposed   model  matches with  the $SgrA^*_{\mathrm{Keck}}$  observational findings.

The organization of the paper is as follows. In section 2, we reconsider the study of rotating and charged black holes with a cloud of strings and quintessence fields  in NC geometry. Section 3 is devoted to the investigation  of the shadows and the energy emission rate of such   NC  rotating and charged black holes using CUDA techniques. Section 4 bridges  the theoretical predictions with observational data from the EHT collaboration through CUDA-based computations. In section 5, we exploit  machine learning methods   to  classify whether the  reduced parameter space of such  NC black holes is consistent with the observations of $M87^*$, Sgr$A^*_{\mathrm{VLTI}}$, and Sgr$A^*_{\mathrm{Keck}}$. Finally, the last section  exposes some concluding remarks.

\section{ NC quintessential rotating   and black holes with a cloud of strings
}

\begin{figure}[h!]
\begin{center}
\includegraphics[width=5cm,height=5cm]{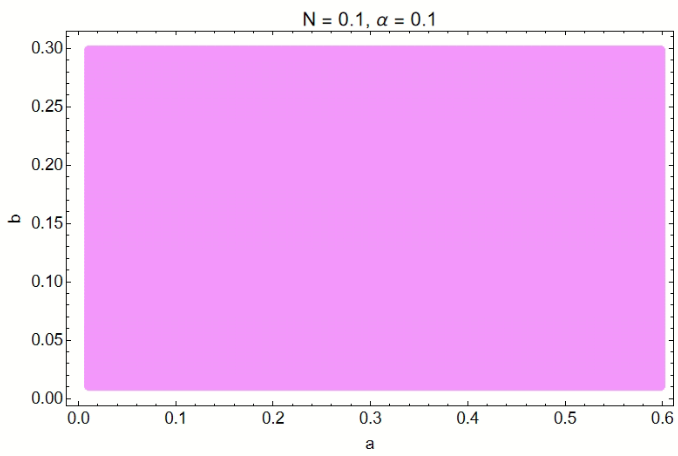}
\includegraphics[width=5cm,height=5cm]{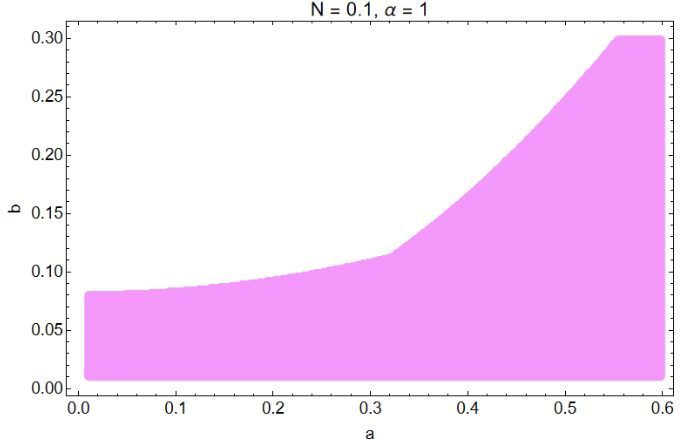}
\includegraphics[width=5cm,height=5cm]{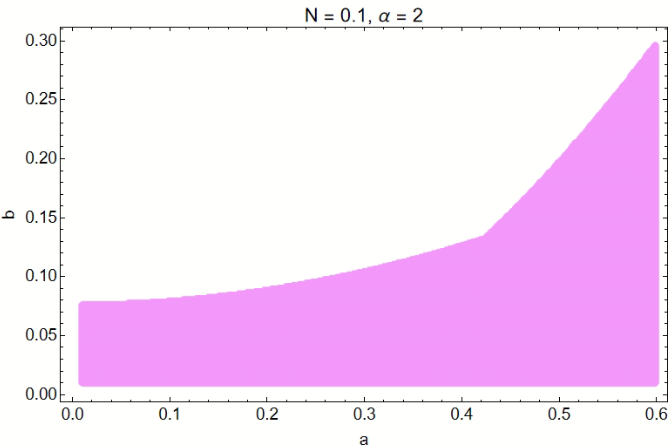}
\end{center}
\caption{Regions in the $(a,b)$ plane where the metric admits at least one real event horizon radius.}
\label{im111}
\end{figure}

In this section, we  reconsider the  study of rotating  and charged  black holes in NC spacetime, characterized by a single parameter $\Theta$. Such NC structures have attracted considerable attention in recent years, particularly in  the investigation of black hole  physical  behaviors \cite{100}.  These studies have been supported by supergravity models where   the NC geometry   naturally arises in string theory  with  D-brane configurations in the presence of a Neveu--Schwarz  antisymmetric $B$-field \cite{101}.   NC spaces have been approached   using physical and  geometrical contributions.   In particular, the gauge theories on such spaces  have been investigated \cite{NC1}.  Moreover, it has been suggested that   NC parameters   can be exploited to remove certain singularities of   complex  geometries including   Calabi-Yau  varieties  with ADE  configurations \cite{NC2, NC3,NC4,NC5}.
 Setting $\hbar = 1$,  the spacetime coordinates  are interpreted as  non-commuting operators satisfying the commutations relations 
\begin{equation}
[x^\mu , x^\nu] = i\,\theta^{\mu\nu},
\end{equation}
where $\theta^{\mu\nu}$ is a constant antisymmetric tensor. In the strong-field limit, it has been shown  that  this tensor is related to the inverse of the stringy field  $B$ considered as a fundamental object in string theory. 
For simplicity reasons, we consider  the reduced form
\begin{equation}
\theta^{\mu\nu} = \Theta\, \epsilon^{\mu\nu},
\end{equation}
where $\epsilon^{\mu\nu}$ is the antisymmetric Levi-Civita tensor and $\Theta$ is a  NC parameter with dimensions of length squared. This provides an economical model involving only one parameter.  The present framework  could be considered as an effective phenomenological description of NC gravitational effects rather than a complete quantum gravity theory \cite{N100}. In this way, the  non-commutativity parameter $\Theta$  can be assumed to provide   leading NC corrections to the black hole geometry \cite{100}. The  generated  solutions are mainly valid in the regime where the NC scale $\sqrt{\Theta}$ remains small compared to the  size of the black hole.  Precisely,  here,  we  investigate  rotating and   charged black hole solutions in such NC backgrounds by  analyzing  the impact of $\Theta$ on their geometric and physical properties including the optical ones. This framework could provide a phenomenological setting that sheds light on the physics of black holes in  non-trivial theories of gravity, notably on D-brane configurations and type II string backgrounds. It has been assumed that  the  dynamics of  such  systems could be described by the Einstein--Hilbert action which can be expressed as follows 
\begin{equation}
S= \int d^{4}x\, \frac{\sqrt{-g}} {16\pi G }\left[ \mathcal{R}+ \mathcal{L}_{m}+\mathcal{L}_{c}\right],
\end{equation}
where $\mathcal{L}_{m}$ and $\mathcal{L}_{c}$ represent the  matter and  the charge contributions, respectively \cite{1}.   To approach the associated behaviors,  one could  consider a static and spherically symmetric spacetime  which reads  as 
\begin{equation}
 ds^2 = -f(r) dt^2 + \frac{dr^2}{f(r)} + r^2 d\theta^2 + r^2 \sin^2 \theta\, d\phi^2,
\end{equation}
where the metric function takes the following form 
\begin{equation}
 f(r) = 1 - \frac{2m(r)}{r} + \frac{q^2(r)}{r^2}.
\end{equation}
The radial  functions  $m(r)$ and $q(r)$  can be  determined  by solving  the Einstein equations of motion which can be  expressed as  follows
\begin{equation}
 G_{\mu\nu} + \Lambda g_{\mu\nu} = 8\pi T_{\mu\nu},
\end{equation}
where the stress energy tensor   can be split as 
\begin{equation}
T_{\mu\nu}= T^m_{\mu\nu}+T^c_{\mu\nu}
\end{equation}
according to the mass and the charge contributions, respectively.  In NC geometry,  such  distributions are modeled by Lorentzian profiles
\begin{equation}
 \rho_M(r)= \frac{M \sqrt{\Theta}}{\pi^{3/2} (r^2 + \Theta)^2}, \qquad   
 \rho_Q(r)= \frac{Q \sqrt{\Theta}}{\pi^{3/2} (r^2 + \Theta)^2},
\end{equation}
where $\Theta$ denotes the NC parameter with dimension $[L^2]$ \cite{2,3}.  In this way, $Q$ represents the black hole charge,  while  $M$ denotes the total mass distributed over a region with a linear size of order $\sqrt{\Theta}$ \cite{4,5}.  The smeared mass and  the charge distribution functions can be expressed as
\begin{equation}
\begin{aligned}
m(r) &= \int_{0}^{r} \rho_{M}(r)\, 4\pi r^{2} \,dr, \\
q(r) &= \int_{0}^{r} \rho_{Q}(r)\, 4\pi r^{2} \,dr.
\end{aligned}
\end{equation}
Roughly,  the  calculations   gives   the following expressions 
\begin{equation}
\begin{aligned}
m(r) &= \frac{2M}{\pi} \arctan\!\left(\frac{r}{\sqrt{\pi \Theta}}\right) 
       - \frac{2M \sqrt{\Theta}}{\sqrt{\pi}} \frac{r}{r^2 + \pi \Theta}, \\
q(r) &= \frac{2Q}{\pi} \arctan\!\left(\frac{r}{\sqrt{\pi \Theta}}\right) 
       - \frac{2Q \sqrt{\Theta}}{\sqrt{\pi}} \frac{r}{r^2 + \pi \Theta}.
\end{aligned}
\end{equation}
Considering the  regime $r \gg \sqrt{\Theta}$,  these expressions reduce to 
\begin{eqnarray}
m(r) &=& M - \frac{4M\sqrt{\Theta}}{\sqrt{\pi}r} + \mathcal{O}\!\left(\frac{\Theta^{3/2}}{r^3}\right),\\
q(r) &= &Q - \frac{4Q\sqrt{\Theta}}{\sqrt{\pi}r} + \mathcal{O}\!\left(\frac{\Theta^{3/2}}{r^3}\right).
\end{eqnarray}
 It  is worth nothing  that the above expansion indicates that the  resulting  metric can be viewed as an approximate  version  in the weak NC regime \cite{N100}. Substituting these expressions into the metric function, one gets  the following non-rotating solutions  in NC geometry 
\begin{equation}
 f(r) = 1 - \frac{2M}{r} 
 + \frac{8M\sqrt{\Theta}}{\sqrt{\pi} r^2}+ \frac{Q^2}{r^2}
 - \frac{8Q^2 \sqrt{\Theta}}{\sqrt{\pi} r^3}+
 + \mathcal{O}(\Theta^{3/2}).
\end{equation}
In addition to the  internal black hole  parameters $M$,  and $Q$,  a set of effective deformation parameters could be implemented  in order to enlarge the associated moduli space.    Moreover, one accounts for rotations by  introducing  the spin parameter $a$ which leads to the following spacetime metric
\begin{equation}
ds^2 = \frac{\sigma(r)}{\Sigma(r)} dt^2
- \frac{2a\,\sigma(r)\sin^2\theta}{\Sigma(r)} dt\, d\phi
+ \frac{\Sigma(r)}{\Delta(r)} dr^2
+ \Sigma(r)\, d\theta^2
+ \frac{r^2 + a^2 + a^2\sigma(r)\sin^2\theta}{\Sigma(r)} \sin^2\theta\, d\phi^2,
\end{equation}
where the metric functions are defined by
\begin{equation}
\Sigma(r) = r^2 + a^2 \cos^2\theta, \qquad
\Delta(r) = r^2 f(r) + a^2, \qquad
\sigma(r) = r^2\bigl(1 - f(r)\bigr).
\end{equation}
 Motivated by several previous works, including the recent studies reported in \cite{9}, the effective metric function incorporates extra  contributions going beyond  the  ordinary black hole configurations. In this present work, we further extend the NC  model by including additional external effects in order to describe more realistic astrophysical and cosmological environments.  Precisely,  we focus mainly  on the impact of the non-commutativity aspect  as well as surrounding matter fields which  could  influence the black hole geometry and its observable properties. These extensions  could be  motivated by the fact that realistic black holes are generally not isolated systems. However, they  are expected to be embedded in non-trivial backgrounds involving both quantum-inspired corrections and large-scale matter distributions.  For such reasons, we consider the following extended  form
\begin{equation}
 f(r) = 1 - \alpha - \frac{2M}{r} + \frac{bM}{r^2} + \frac{Q^2}{r^2}
 - \frac{bQ^2}{r^3} - \frac{N}{r^{3w+1}}
\end{equation}
where   one has  introduced a  new   non-commutativity  parameter form   which reads as 
\begin{equation}
b = \frac{8\sqrt{\Theta}}{\sqrt{\pi}}.
\end{equation}
The terms $bM/r^2$ and $-bQ^2/r^3$ arise as perturbative NC corrections in the large-radius regime, consistent with  the Gaussian-smeared source models  \cite{10,11}.  Other external parameters   have been included such as   the string cloud parameter $\alpha$ and the  quintessence   ones.   In particular,  $N$ and $w$  represent  the dark section field  contributions \cite{Q1}.    A close inspection  demonstrates  that the metric functions   generate  a     black  hole  moduli space $ {\cal M}_{bh}$  which can  be spit as 
\begin{equation}
{\cal M}_{bh}=  {\cal M}_{int} \times {\cal M}_{ext}
\end{equation}
where the first factor controls  the internal parameters 
 \begin{equation}
 {\cal M}_{int}=\{ M,Q,a\}
\end{equation}
while the second one represents the external ones 
 \begin{equation}
 {\cal M}_{ext}=\{ b, \alpha, N, w \}.
\end{equation}
A careful examination  reveals that  one can consider  certain special  regions by fixing the value of $w$.   For simplicity  reasons, we set  $w=-\frac{2}{3}$. After  fixing certain regions in the black hole moduli space, we  examine  the existence of horizons using a numerical approach implemented in CUDA, which enables efficient parallel computations on GPU architectures.   It is denoted  that CUDA is a general-purpose computing platform and programming models exploiting  the parallel computing capabilities of NVIDIA GPUs. In this  method, the GPU's architecture enables the distribution of computational tasks across a large number of streaming multiprocessors (SMs),  which  ensures  a   highly efficient parallel processing. Moreover,  the modern GPU's provide advanced optimization techniques and tools  enhancing the  computational performance. With each new generation, improvements in the CUDA core architecture lead to increased efficiency and reduced computation time. In  this context,  the CUDA has proven to be a powerful tool in the  black hole physics, enabling efficient GPU-accelerated simulations \cite{f34,s1,s2}. This significantly not only  reduces computational cost but also improves  the numerical efficiency. Besides,  it can  facilitate the exploration of theoretical predictions in strong gravity regimes \cite{s3,s4}. To verify the numerical reliability of the present analysis, additional  tests  have been performed by changing the integration step and the grid resolution used in the simulations. The obtained results show stable convergence with only small numerical deviations, indicating a satisfactory numerical precision in the explored parameter space \cite{num1,num2,num3}. In fact, the CUDA-based GPU implementation considerably reduces the computation time compared to standard CPU calculations, especially for large parameter scans and high-resolution numerical simulations \cite{cuda1,cuda2}.  This  acceleration makes the present framework suitable for extensive computations involving many parameter configurations and geodesic evaluations \cite{bhnum1,bhnum2}. Roughly, the horizons are identified by   vanishing the inverse of  the radial component of  the  metric  function. To probe their existence, we employ a numerical method in which the parameters $Q$, $\alpha$, and $N$  remain  constant, while the parameter $b$ varies from $0$ to $0.3$ in increments of $0.1$. The parameter $a$ can be   adjusted over the same range using the same step size. For each pair $(a,b)$, the horizon equation is  solved numerically  in order  determine whether there exists at least one real root corresponding to a physically meaningful horizon in  black hole physics. This systematic procedure provides a way to identify the regions of  the  reduced  back hole   moduli space in which these solutions are acceptable.  Fig.~(\ref{im111})  illustrates   such behaviors by  displaying  the  regions in the $(a,b)$ plane where at least one real horizon exists, for different values of $N$ and $\alpha$.

This figure shows  that increasing the parameter $\alpha$ reduces the regions  of  the parameter space  which   supports the existence of  physical horizons. Based on these results, the forthcoming optical analysis will be conducted within the parameter  subsets  which correspond  to configurations admitting at least one real horizon in the associated black hole physics.

\section{ NC black hole shadows and energy emission rate using CUDA computations}
\begin{figure}[!h]
\centering
\includegraphics[width=8cm,height=7cm]{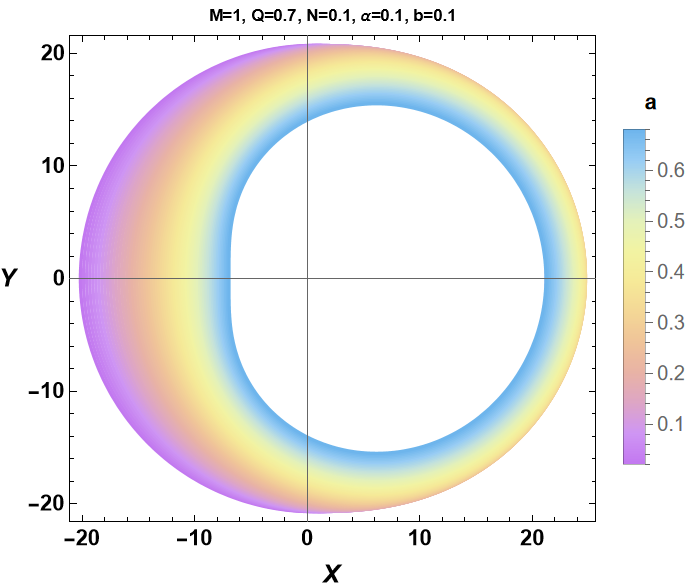}
\includegraphics[width=8cm,height=7cm]{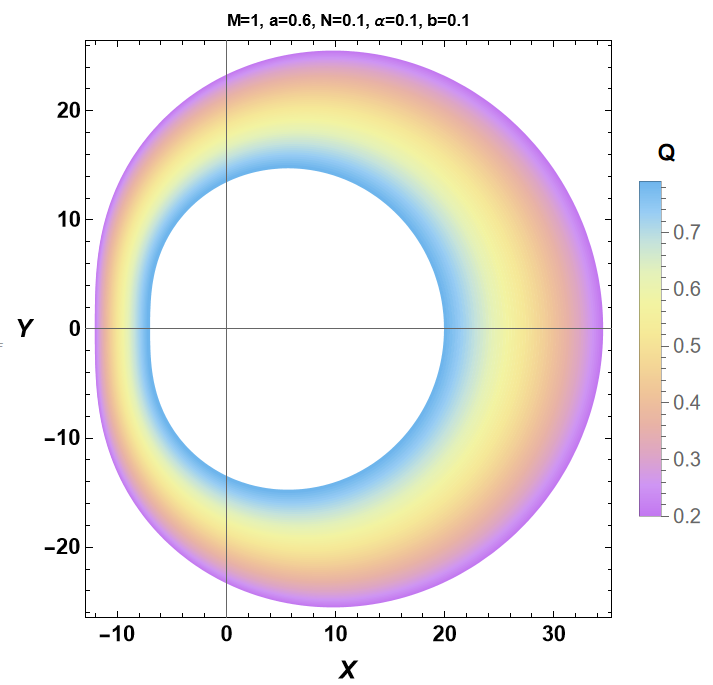}
\caption{\textit{\protect\footnotesize Effect of internal parameters on the shadow behavior.}}
\label{im44}
\end{figure}

In this section, we  investigate   certain optical properties  of   rotating and  charged  NC black holes with a cloud of strings and quintessence  in NC  geometry by  applying CUDA techniques.  Precisely, we analyze the effect  the involved  parameters on both the black hole shadow and the energy emission rate. To elaborate this analysis,  we employ a CUDA-based numerical code enabling high-performance parallel computations. It is  denoted that  the shadow of a black hole is defined as the apparent dark region bounded by the critical curve observed when light rays approach unstable circular photon orbits before being captured or scattered \cite{HerdeiroRadu2014, CunhaHerdeiro2018,230,231}. This  optical  aspect  can be  examined by approaching  the  null geodesics in the black hole spacetime. On the other hand, the energy emission rate is directly  linked  to the absorption cross section of the black hole  which provides significant  information about its thermodynamic and radiative properties. To investigate the optical properties of such  rotating  and charged  NC  black holes,  roughly, we  establish the associated  equations of motion by  implementing the CUDA techniques in  the Hamilton–Jacobi formalism via   the separation of variables according to  the Carter method \cite{26}. For the above  NC black hole solutions, the resulting equations of motion can be expressed  as follows
\begin{align}
\Sigma \dot{t}& =\frac{r^{2}+a^{2}}{\Delta }\left[ E\left(
r^{2}+a^{2}\right) -aL\right] +a\left[ L-aE\sin ^{2}\theta \right] \\
(\Sigma \dot{r})^{2}& =\mathcal{R}(r) \\
(\Sigma \dot{\theta})^{2}& =\Theta (\theta ) \\
\Sigma \dot{\phi}& =\left[ L\csc ^{2}\theta -aE\right] +\frac{a}{\Delta }%
\left[ E\left( r^{2}+a^{2}\right) -aL\right] ,
\end{align}%
where $E$ and $L$ are the energy and the angular momentum of the light rays,
respectively. $\mathcal{R}(r)$ and $\Theta (\theta )$ are radial  functions  which read as 
\begin{align}
\mathcal{R}(r)& =\left[ E\left( r^{2}+a^{2}\right) -aL\right] ^{2}-\Delta %
\left[ \mathcal{C}+\left( L-aE\right) ^{2}\right] , \\
\Theta (\theta )& =\mathcal{C}-\left( L\csc \theta -aE\sin \theta \right)
^{2}+\left( L-aE\right) ^{2},
\end{align}%
where $\mathcal{C}$ is a  Carter separation parameter. Solving the unstable
circular orbit equations, one needs two impact parameters  being  given by 
\begin{align}
\xi &= \frac{r^{2}\left[16a^{2}\Delta(r) + 8r\,\Delta(r)\,\Delta'(r)^{2} - r^{2}\,\Delta'(r)^{2}\right]}{a^{2}\,\Delta'(r)^{2}} \bigg|_{r = r_{0}}, \label{3.5} \\[6pt]
\Xi &= \frac{(r^{2} + a^{2})\,\Delta'(r) - 4r\,\Delta(r)}{a\,\Delta'(r)} \bigg|_{r = r_{0}}. \label{3.6}
\end{align}
In the case of rotating black holes with a cloud of strings and quintessence fields  in NC geometry, the apparent shadow, at spatial infinity,  can be  illustrated   using the celestial coordinates $(X,Y)$  which are  expressed as follows
\begin{align*}
X& =\lim_{r_{\text{ob}}\rightarrow +\infty }\left( -r_{\text{ob}}^{2}\sin
\theta _{\text{ob}}\frac{d\phi }{dr}\right) \\[4pt]
Y& =\lim_{r_{\text{ob}}\rightarrow +\infty }\left( r_{\text{ob}}^{2}\frac{%
d\theta }{dr}\right) ,
\end{align*}%
where $r_{\text{ob}}$ denotes the distance between the observer and the black hole. The quantity $\theta_{\text{ob}}$ represents the inclination angle between the observer line of the sight and the rotation axis of the black hole. To investigate the influence of each parameter on the black hole shadows,  we employ a numerical approach to provide the corresponding   graphical representation   for acceptable ranges  of the involved black hole parameters. More specifically, a CUDA-based parallel computing program  has been used to accelerate the calculations~\cite{43,44}, which allows for the efficient determination of the  shadow boundaries for different parameter variations. For each set of shadow curves, all parameters are held constant except for the parameter of interest, which is varied in increments of $0.01$. Considering the special case $w = -\frac{2}{3}$, we solve Eqs.~(\ref{3.5}) and (\ref{3.6}) and substitute the resulting expressions into the shadow equation. This procedure permits  to accurately evaluate the influence of each parameter on the geometric deformation of the shadow, affecting both its size and shape. In Fig.~(\ref{im44}), we  illustrate  the black hole shadows by analyzing the effects of the charge $Q$ and the rotation parameter $a$, by  taking  $b = 0.1$ and $\alpha = 0.1$.

As shown in the figure, the rotation parameter $a$ exhibits behaviors similar to that in standard black hole solutions. Indeed, it slightly reduces the size of the shadow and deforms the shape into a D-like structure. This confirms that $a$ keeps  its role as a deformation parameter in these configurations.  Increasing $Q$, however, the size of the shadow decreases without affecting the shape, which is consistent with the behavior observed in ordinary charged black holes. However, in the present model, the range of variation in the shadow size extends up to values of approximately $45$, unlike ordinary charged black holes, where the radius generally does not exceed $25$. This difference can be attributed to the additional terms in the metric, in which the parameter $b$ couples with the charge $Q$. Then, the effect of the string cloud parameter $\alpha$ and the quintessence parameter $N$ are examined by varying one while keeping the other one  fixed, thereby isolating their respective contributions. For $b = 0.1$,  the variation of $\alpha$ and $N$  increases the overall size of the black hole shadow.

\begin{figure}[!h]
\centering
\includegraphics[width=8cm,height=7cm]{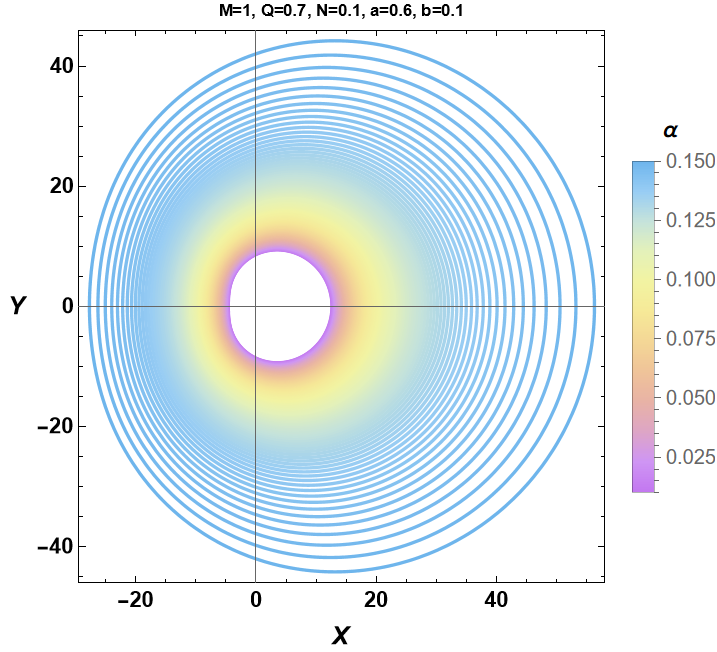}
\includegraphics[width=8cm,height=7cm]{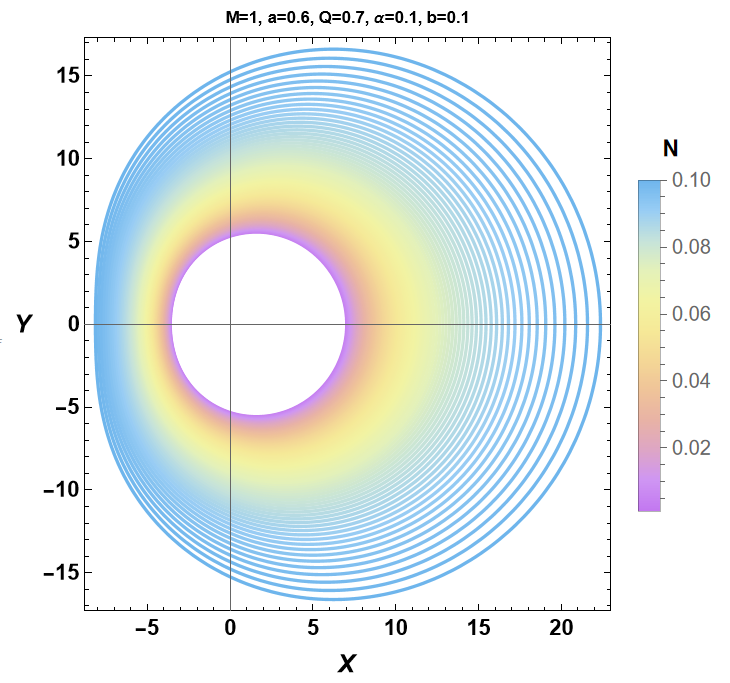}
\caption{\textit{\protect\footnotesize Effect of $\alpha$ (left panel) and $N$ (right panel) on the black hole shadow behavior.}}
\label{im4}
\end{figure}

As illustrated in Fig.~(\ref{im4}), the shadow exhibits  a D-like deformation in the regime of large $N$ and small $\alpha$. This characteristic gradually reduces as $\alpha$ increases or $N$ decreases, which indicates that these two parameters have a similar influence on the distortion of the shadow curves.

\begin{figure}[!h]
\centering
\includegraphics[width=8cm,height=7cm]{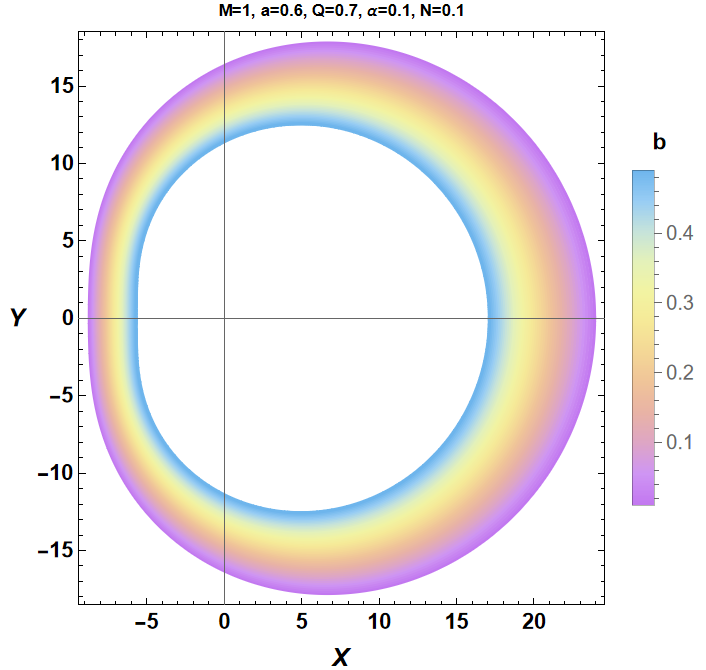}
\includegraphics[width=8cm,height=7cm]{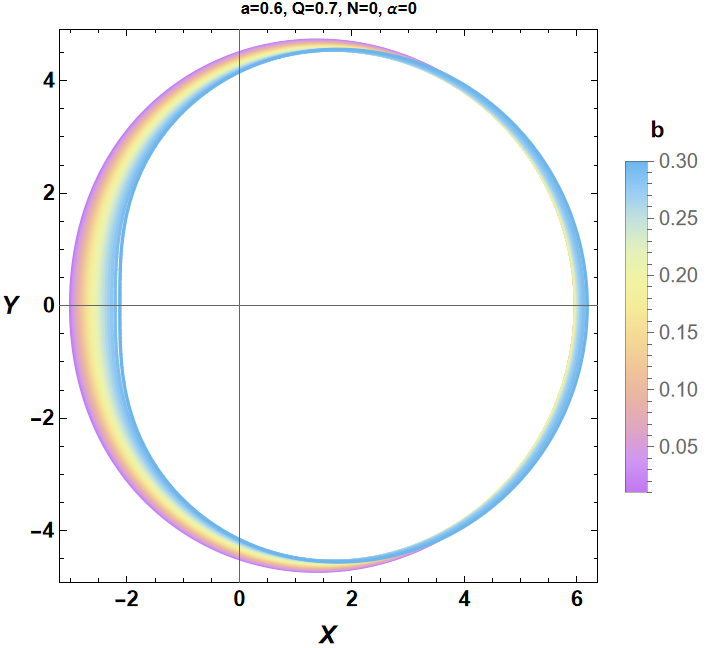}
\includegraphics[width=8cm,height=7cm]{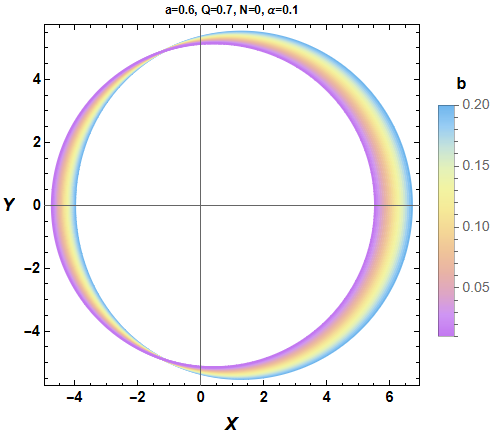}
\includegraphics[width=7.9cm,height=6.8cm]{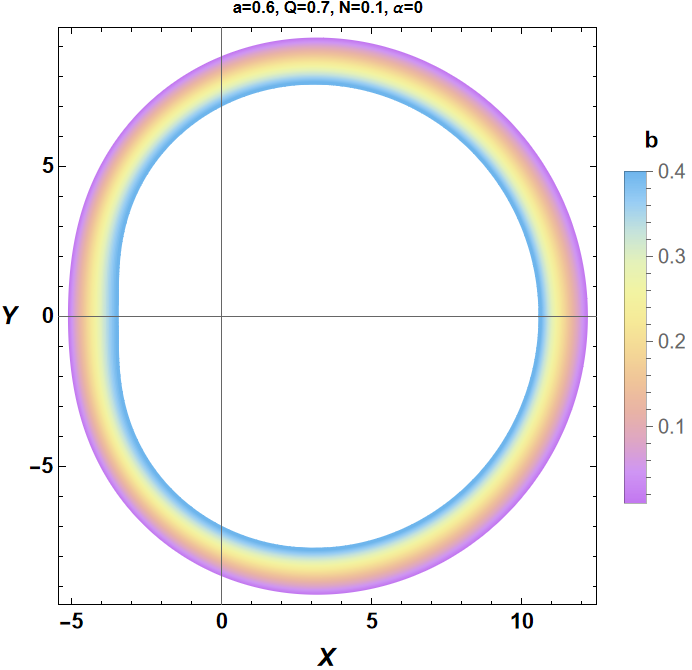}
\caption{\textit{\protect\footnotesize Effect of the parameter $b$ on the shadow behavior.}}
\label{im45}
\end{figure}

We now examine the effect  of the  NC parameter $b$ on  the shadow behaviors.  As shown in Fig.~(\ref{im45}),  such a parameter   controls both the size and the global shape of the shadow. Increasing $b$ leads to a larger and more extended shadows, whereas smaller values of $b$ yield a more compact and nearly circular geometric configurations.

In contrast, the parameters $N$ and $\alpha$ have less effects,  mainly changing the shape of the shadow boundary rather than its size. Consequently, $b$ sets the characteristic size of the shadow, while $N$ and $\alpha$ control deviations from the  circularity and the resulting distortions of its shape.

To  go beyond  the  discussion of the shadow properties of black holes with a cloud of strings and the  quintessence in NC geometry, we employ a CUDA-based numerical method to evaluate the energy emission rate associated with rotating black holes in this non-trivial  framework. By leveraging the parallelism of GPUs,  this approach provides efficient and high-performance computations of  the effective absorption cross sections over a wide range of parameters. For a distant observer, the absorption cross section at very high energies asymptotically  can reach a geometric optics limit, which is directly related to the size of the black hole shadow. At intermediate energy regimes, the effective absorption cross section exhibits oscillations around a constant value, denoted by $\sigma_{\text{lim}}$. It has been shown that this limiting value coincides with the geometric cross section of the photon sphere, which is determined by the properties of null geodesics \cite{27,28,29}.  Since the shadow encodes the optical appearance of the black hole, it can be identified with this limiting cross section. Therefore, $\sigma_{\text{lim}}$ can be approximated as follows 
\begin{equation}
\sigma_{\text{lim}} \simeq \pi R_{s}^{2},
\end{equation}
where $R_s$ represents the radius of the black hole shadow. In this context, the differential energy emission rate can be written as
\begin{equation}
\frac{d^{2}E(\omega)}{d\omega \, dt} = \frac{2\pi^{3} R_{s}^{2}}{%
e^{\omega/T_{H}}-1} \, \omega^{3},
\end{equation}
where $T_{H}$  denotes  the Hawking temperature of the black hole and $\omega$  represents the emission frequency. This relation establishes a direct connection between the thermodynamic properties of the black hole and its optical features. In particular, it  generates  a useful framework for probing spacetime parameters through observational signatures. For the rotating metric solutions, the Hawking temperature of such black holes reads as 
\begin{equation}
T_{H}=\frac{2r(1-\alpha)-2M+\frac{bQ^2}{r^2}-\frac{N(1-w)}{r^{2w}}}{4 \pi \left(a^{2}+r^{2}\right)}.
\end{equation}
To evaluate the energy emission rate for different black hole parameter values, we perform numerical simulations using a CUDA-based program. The corresponding code first computes the maximal shadow radius from the obtained optical data. Then, the horizon radius is determined by solving $\Delta(r_h)=0$,. Moreover, the result is substituted into the Hawking temperature expression.  Taking the  specific case $w = -\frac{2}{3}$, we assess the influence of each parameter by varying the parameter of interest in steps of $0.01$ while keeping all other parameters fixed. The resulting data are then used to generate the energy emission rate plots.

\begin{figure}[!ht]
\begin{center}
\centering
\begin{tabbing}
			\centering
			\hspace{0.5cm}\=\kill
					\includegraphics[scale=0.35]{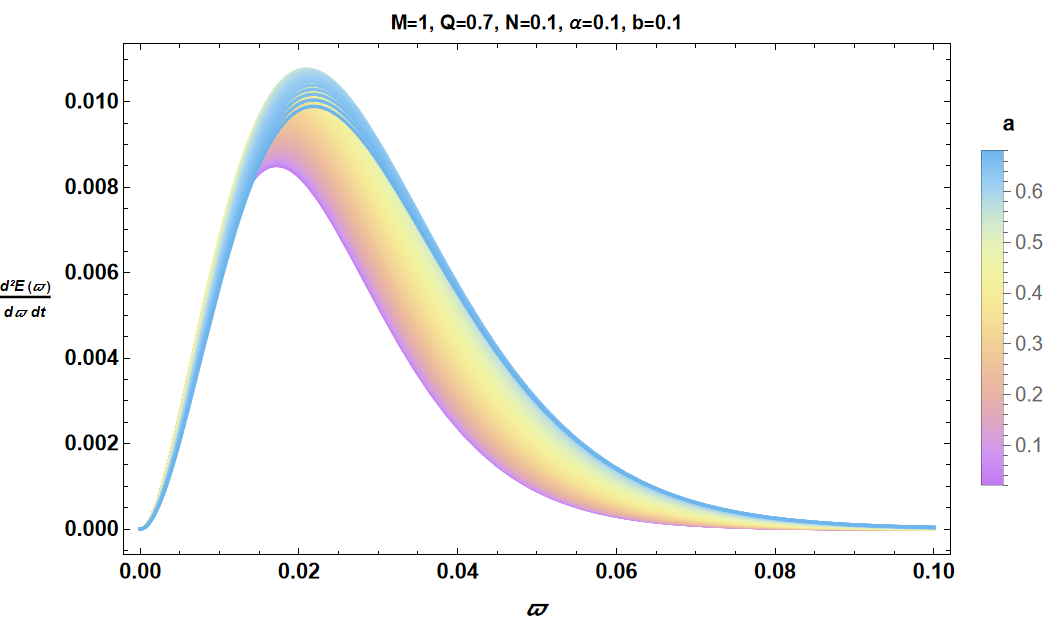}\hspace{0.05cm}	\includegraphics[scale=0.35]{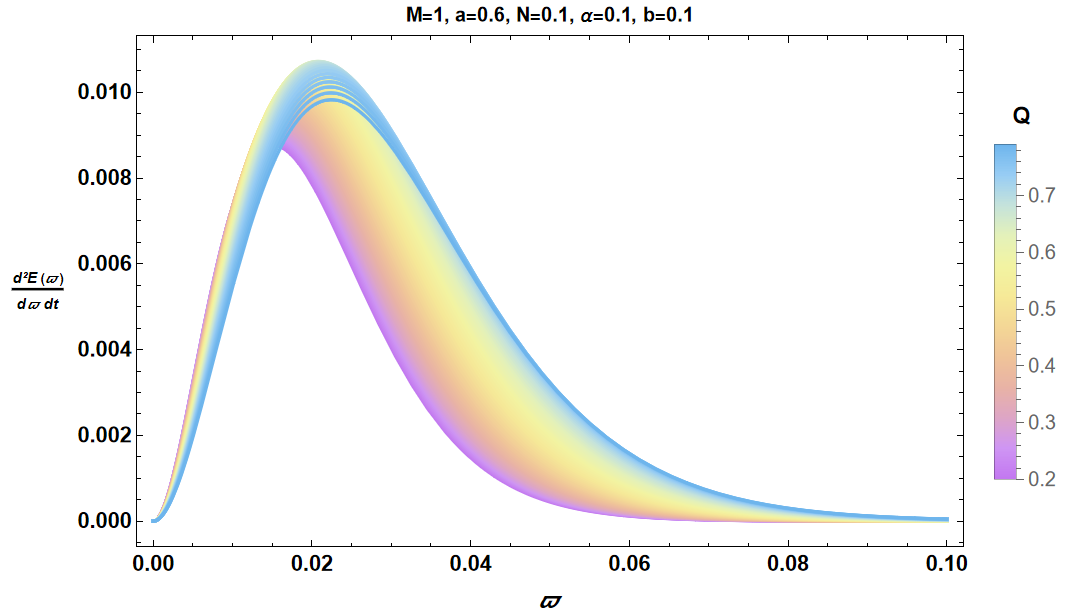}\\ 
	 \end{tabbing}
\end{center}
\caption{ \textit{\protect\footnotesize Variation of the energy emission
rate as a function of the emission frequency for different values of $a$ and 
$Q$. }}
\label{ER}
\end{figure}

In Fig.~(\ref{ER}), we present the variation of the energy emission rate as a function of the emission frequency. The figure consists of two panels. The first  one  illustrates the effect of the electric charge, while the second  one shows the effect of the rotation parameter. This comparison unveil  the role of  such  parameters in the emission process graphical representation.

It  has been  observed from both panels that increasing either the charge $Q$ or the rotation parameter $a$ leads to an initial enhancement of the energy emission rate up to a critical value, after which the emission rate decreases. Thus, both $Q$ and $a$ manifest  a non-monotonic behavior, acting as amplifying factors at low values and as suppressing effects beyond their respective critical thresholds.

\begin{figure}[!ht]
\begin{center}
\centering
\begin{tabbing}
			\centering
			\hspace{0.5cm}\=\kill
					
	\includegraphics[scale=0.35]{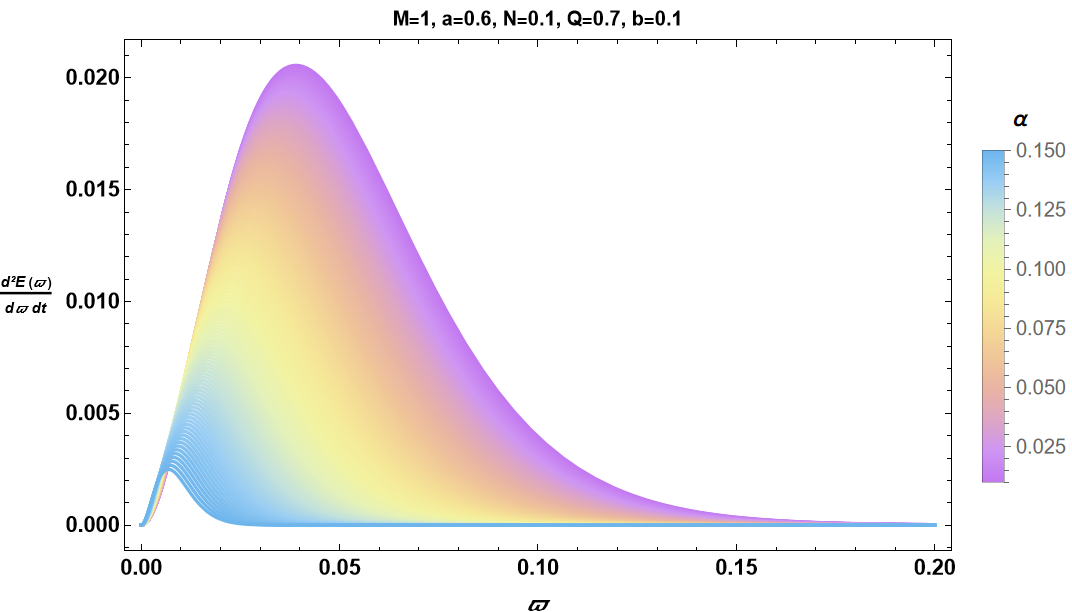} 
	\hspace{0.05cm}		\includegraphics[scale=0.35]{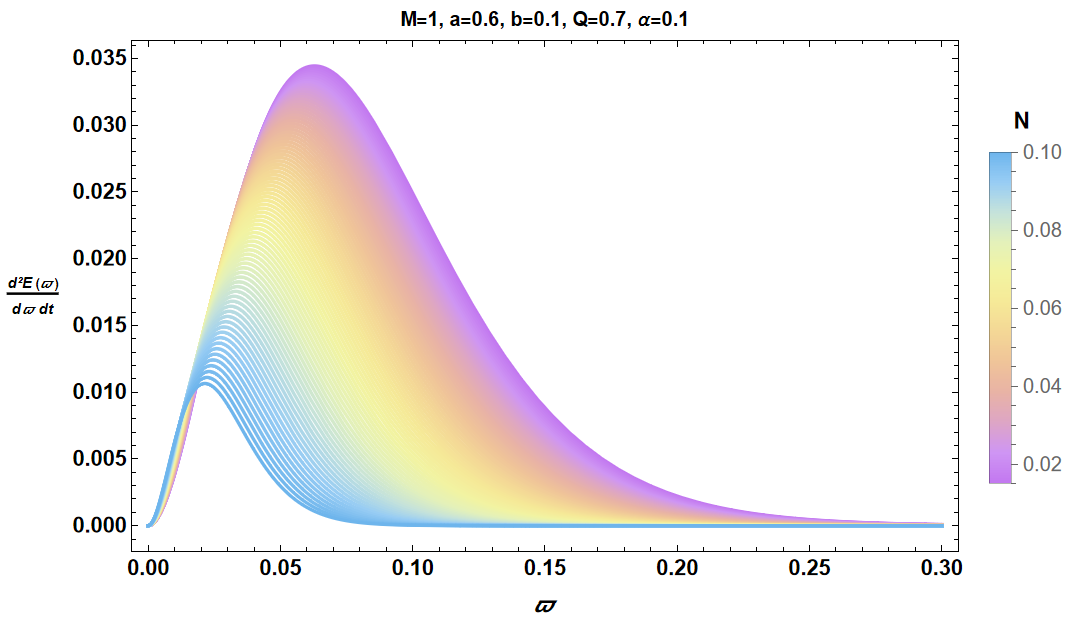}\\  
	 \end{tabbing}
\end{center}
\caption{ \textit{\protect\footnotesize Variation of the energy emission
rate as a function of the emission frequency for different values of $%
\protect\eta$. }}
\label{ER1}
\end{figure}

Fig.~(\ref{ER1}) shows the effects of the parameters $\alpha$ and $N$ on the energy emission rate. Both $\alpha$ and $N$ produce a clear suppression of the energy emission rate. This behavior indicates that increasing either $\alpha$ or $N$ reduces the efficiency of the radiation process in this regime. Thus, these results clearly show the attenuating effect of these two parameters on the emission spectrum.

\begin{figure}[!ht]
\centering
\includegraphics[scale=0.35]{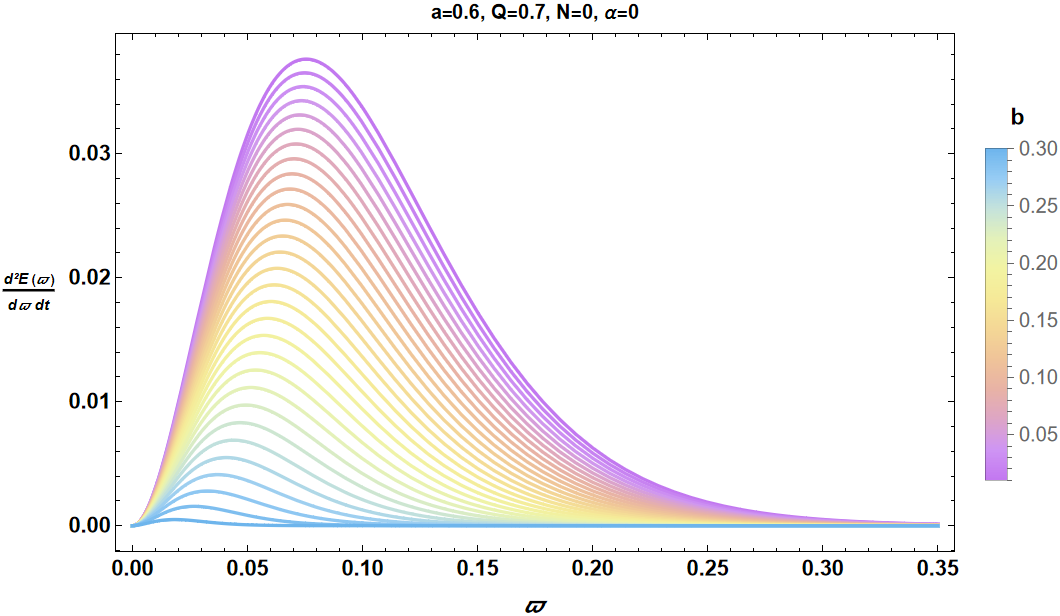}
\includegraphics[scale=0.35]{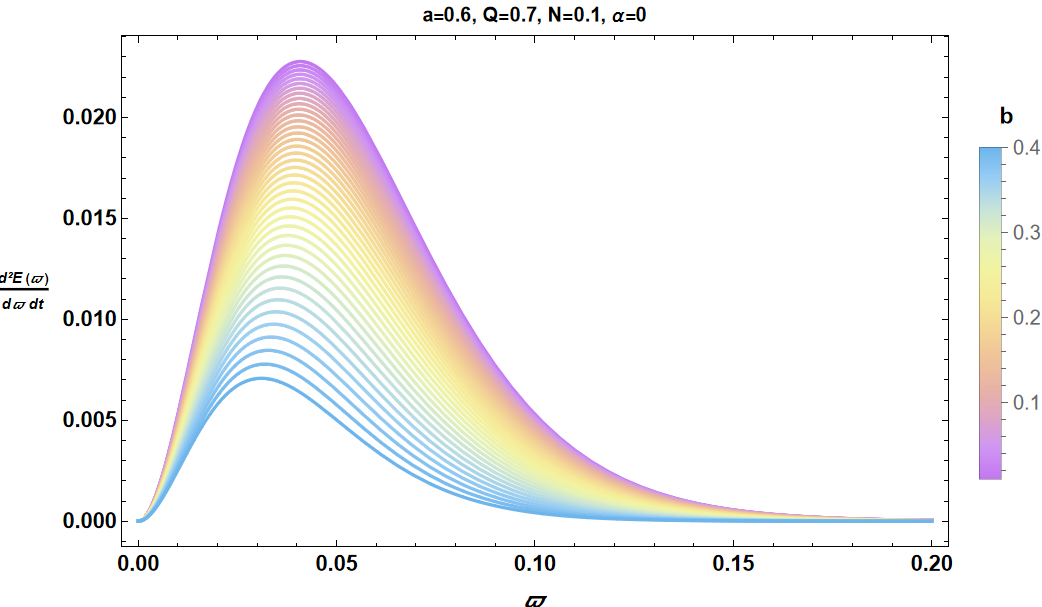}
\includegraphics[scale=0.35]{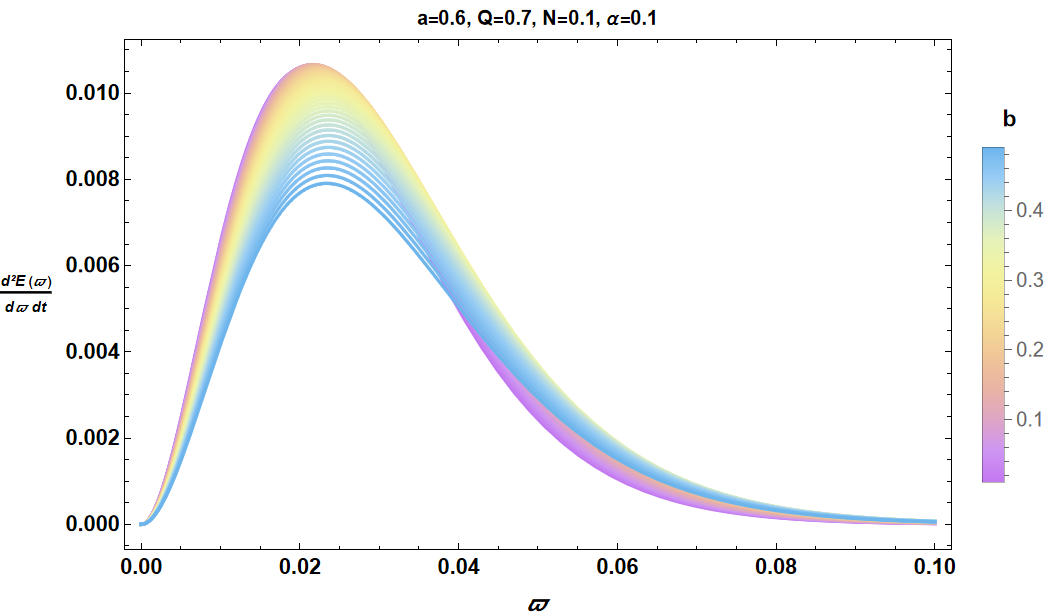}
\includegraphics[scale=0.35]{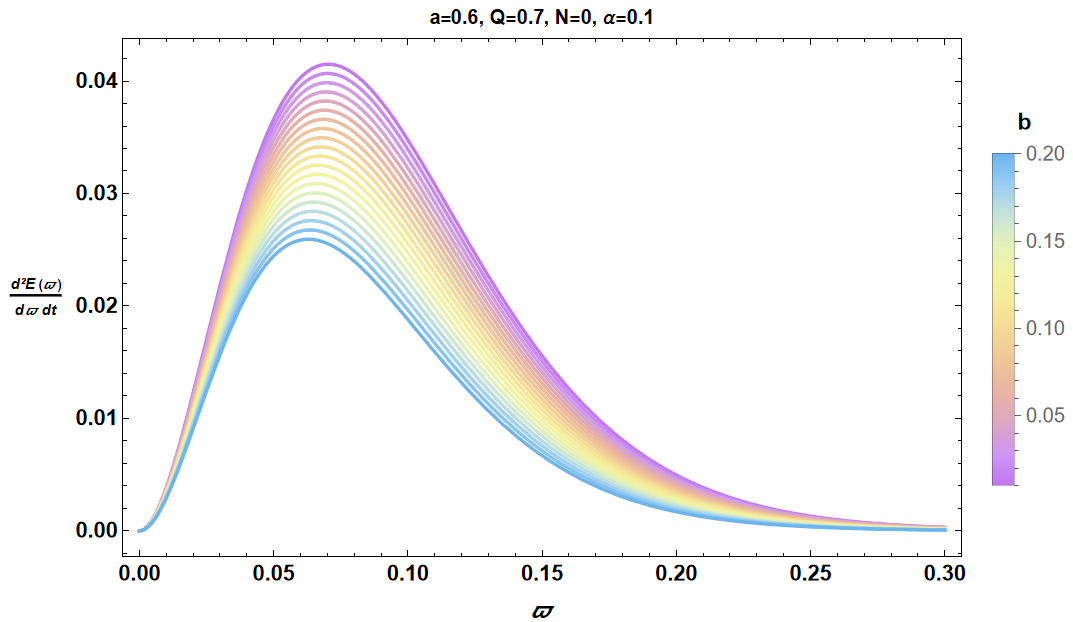}
\caption{ \textit{\protect\footnotesize Variation of the energy emission
rate as a function of the emission frequency for different values of $b$ .}}
\label{Eb}
\end{figure}

From the four panels in Fig.~(\ref{Eb}), we analyze the effect of the NC parameter \(b\) on the energy emission rate for different values of \(N\) and \(\alpha\). In all cases, the general behavior follows a similar  behavior. Indeed, the energy emission rate increases to a maximum value and then decreases as the frequency parameter increases. It is clear that an increase in \(b\) results in a reduction in the peak emission rate as well as a decrease in the overall amplitude of the curve. This means that higher values of \(b\) reduce the energy emission process and make the radiation less intense. This behavior is consistent in all subgraphs, independent of the values selected for  $N$  and  $\alpha$. Moreover, it indicates that $b$ has a reducing effect on the emission spectrum.

\section{Constraints on NC rotating  black hole parameters from EHT observations using  CUDA techniques}

In order to establish a bridge  between the  theoretical predictions and  the observational data, this section  provides an analysis of the shadow cast by rotating NC  quintessence Reissner–Nordström black holes with a cloud of strings, in comparison with observational results reported by EHT  empirical  findings.  Precisely,  we use the observational data of the M87*  and Sgr A* black holes to constrain the parameters of these black hole models~\cite{E1,E2,E3}.  In the present work,  the numerical analysis  has been  carried out using a CUDA-based code developed by NVIDIA, which exploits GPU parallel computing to significantly accelerate the computations required for determining the black hole shadows. In practice, the constraints can be obtained by using the fractional deviation from the Schwarzschild black hole shadow diameter, defined as
\begin{equation}
{\ d} = \frac{R_s}{r_{sh}}-1,
\end{equation}
where $R_s$ denotes the shadow radius and $r_{\text{sh}}$ represents the Schwarzschild radius. The dimensionless ratio $R_s/M$ serves as a primordial observable  to  confront the theoretical models with observational data.  The  $1\sigma$  and  $2\sigma$  confidence intervals derived from the EHT observations are displayed in Table ~\ref{t1}.

\begin{table}[h!]
\centering
\begin{tabular}{|c|c|c|c|}
\hline
\textbf{Black hole} & \textbf{Deviation ($d$)} & \textbf{1-$\sigma$ bounds}
& \textbf{2-$\sigma$ bounds} \\ \hline
M87$^*$ (EHT) & $-0.01^{+0.17}_{-0.17}$ & $4.26 \leq \frac{R_s}{M} \leq 6.03$
& $3.38 \leq \frac{R_s}{M} \leq 6.91$ \\ \hline
Sgr~A$^*$ (EHT$_{\text{VLTI}}$) & $-0.08^{+0.09}_{-0.09}$ & $4.31 \leq \frac{%
R_s}{M} \leq 5.25$ & $3.85 \leq \frac{R_s}{M} \leq 5.72$ \\ \hline
Sgr~A$^*$ (EHT$_{\text{Keck}}$) & $-0.04^{+0.09}_{-0.10}$ & $4.47 \leq \frac{%
R_s}{M} \leq 5.46$ & $3.95 \leq \frac{R_s}{M} \leq 5.92$ \\ \hline
\end{tabular}%
\caption{ \textit{\protect\footnotesize  Fractional deviations and
corresponding bounds for M87$^*$ and Sgr~A$^*$ black holes.}}
\label{t1}
\end{table}

\begin{figure}[h!]
\centering

\includegraphics[scale=0.295]{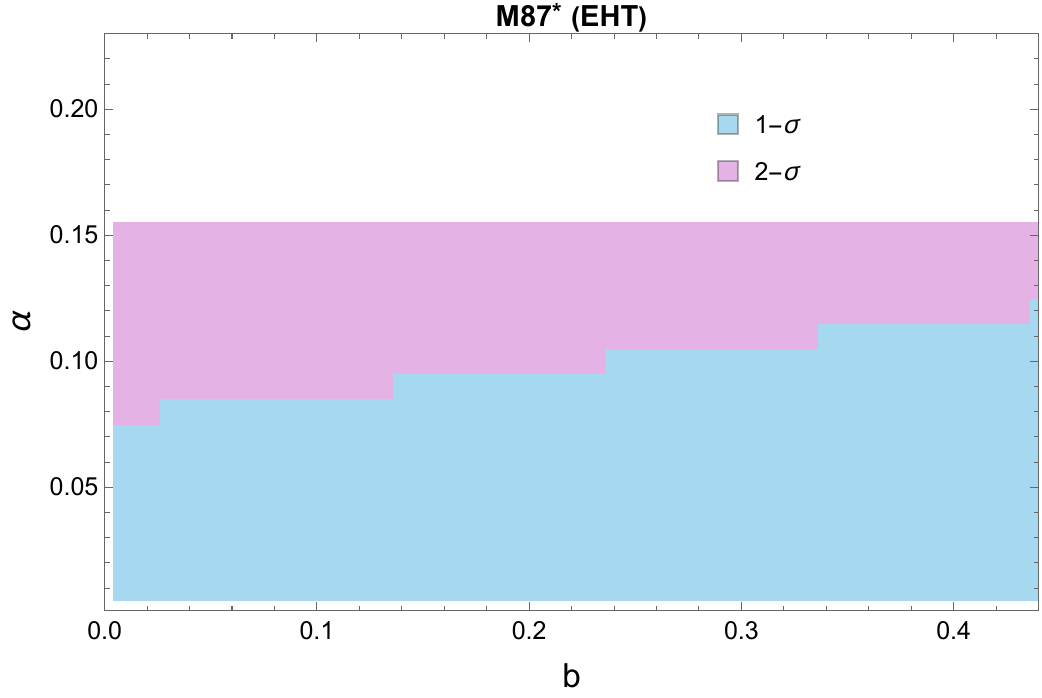}
\includegraphics[scale=0.295]{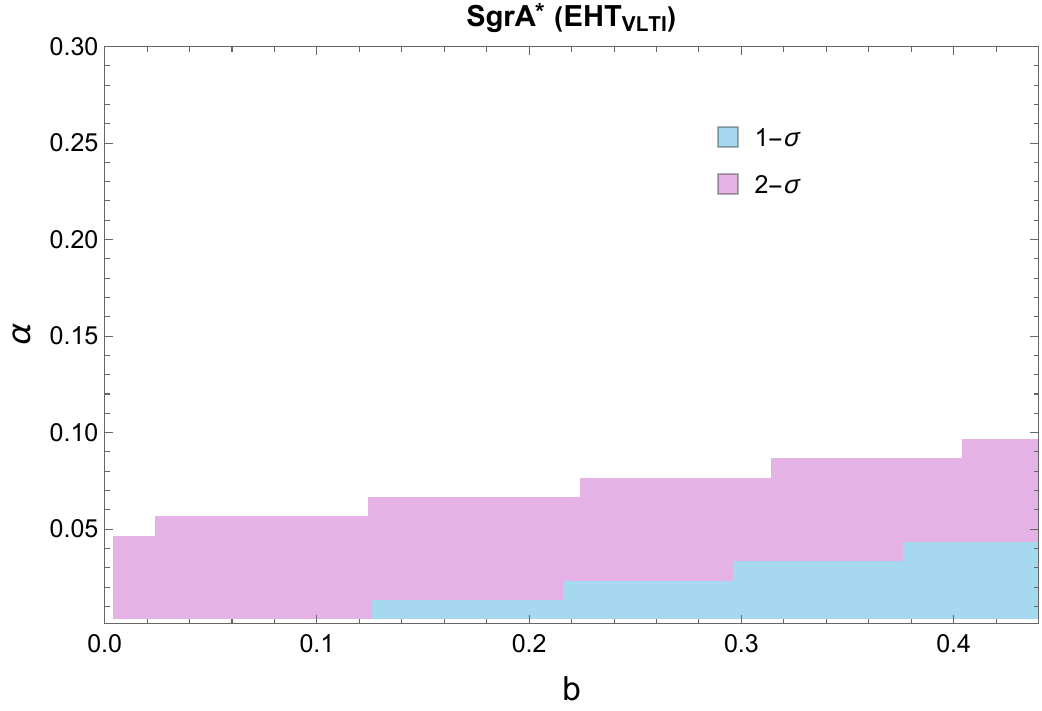}
\includegraphics[scale=0.295]{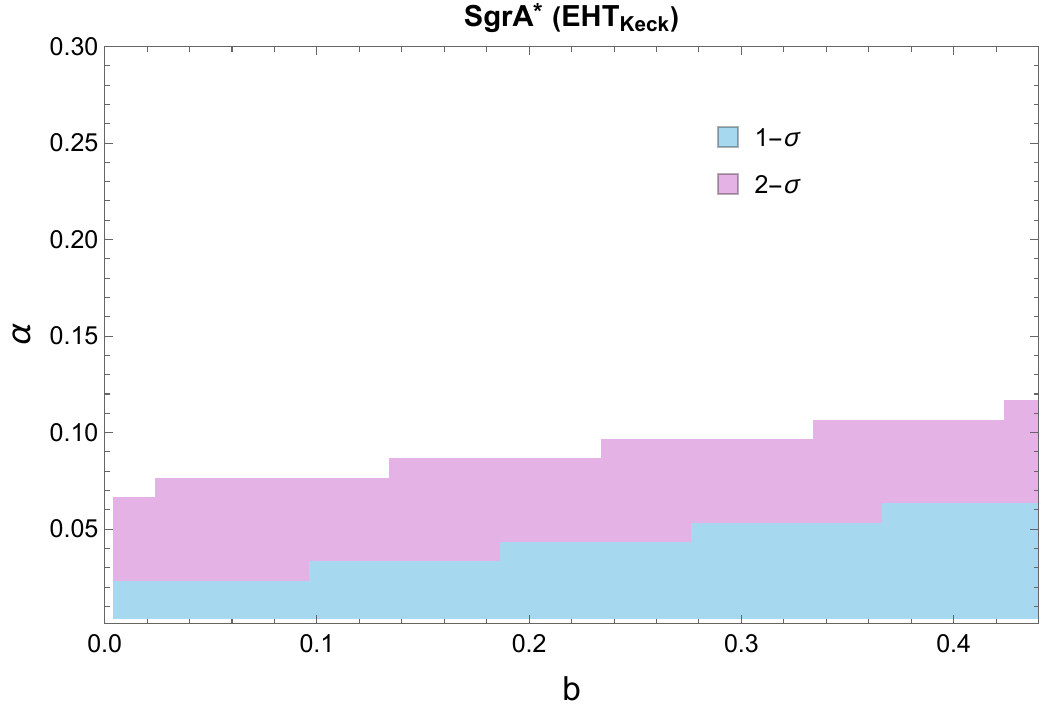}

\vspace{0.2cm}

\includegraphics[scale=0.295]{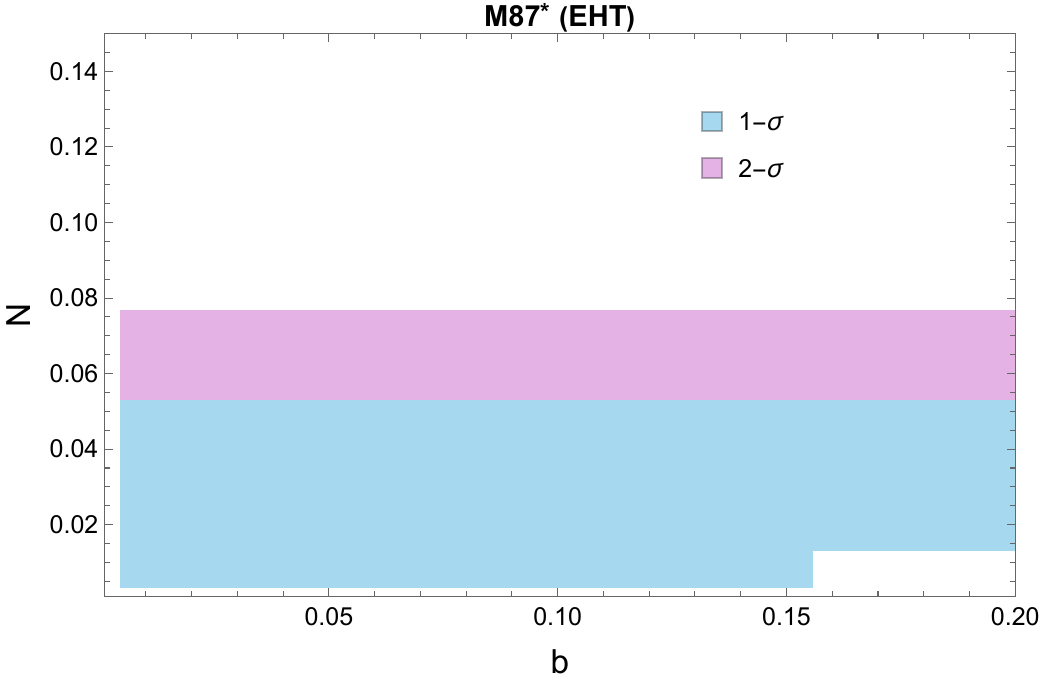}
\includegraphics[scale=0.295]{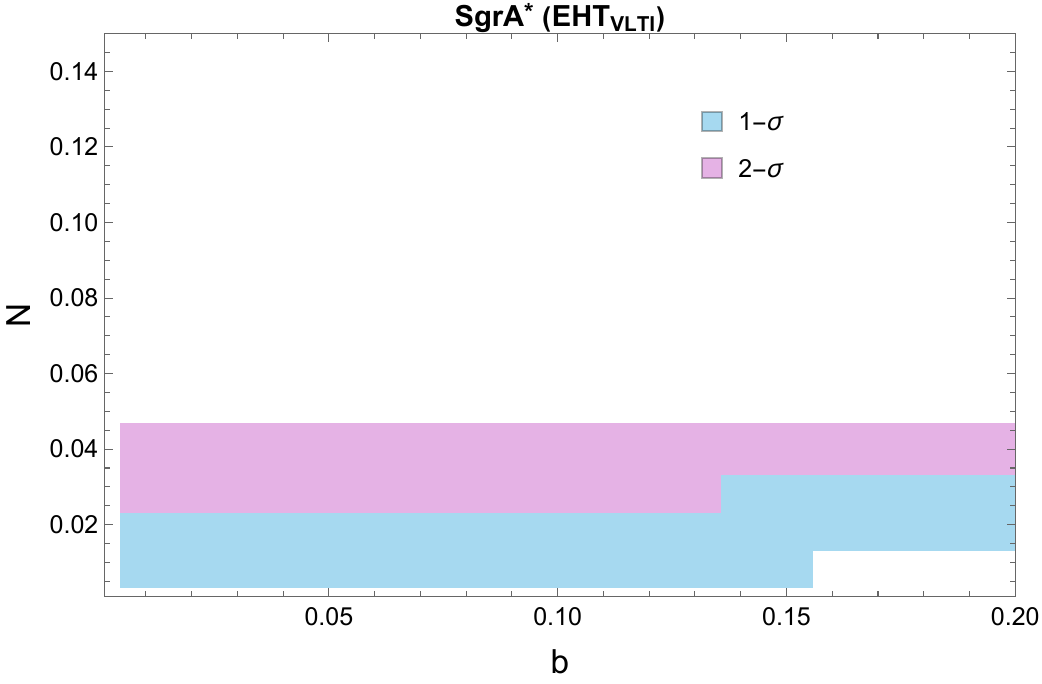}
\includegraphics[scale=0.295]{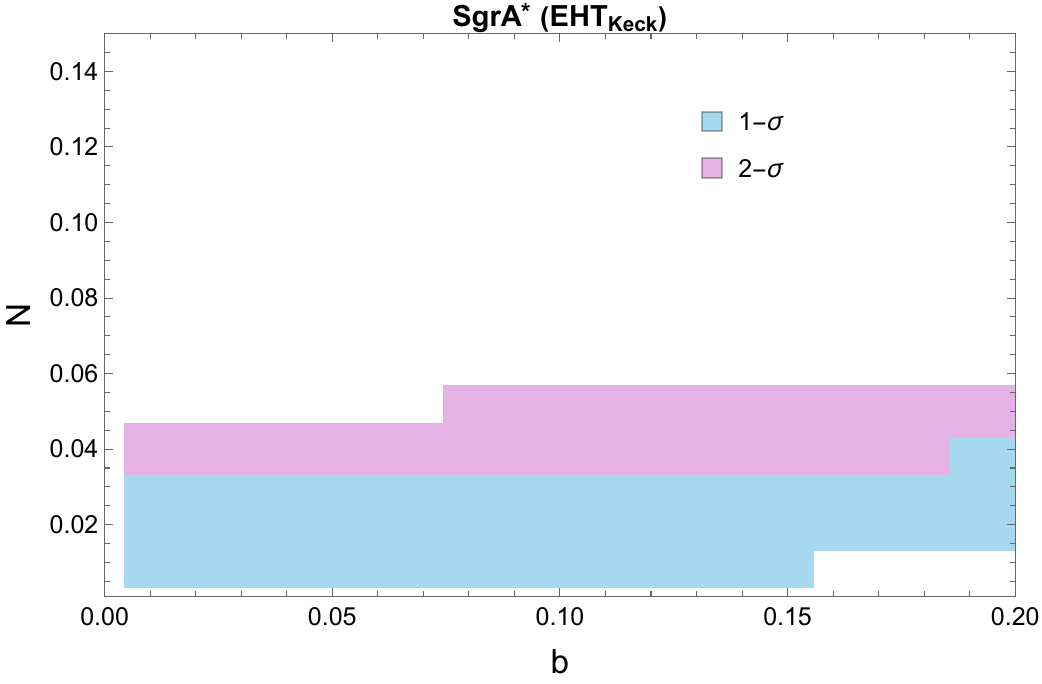}

\vspace{0.2cm}

\includegraphics[scale=0.295]{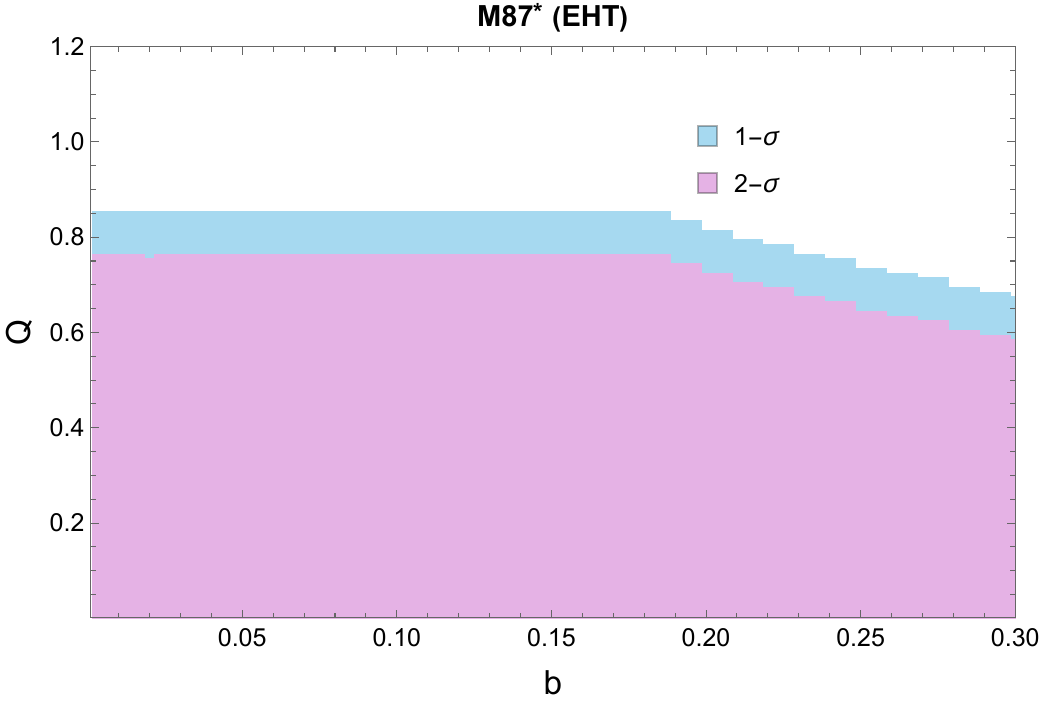}
\includegraphics[scale=0.295]{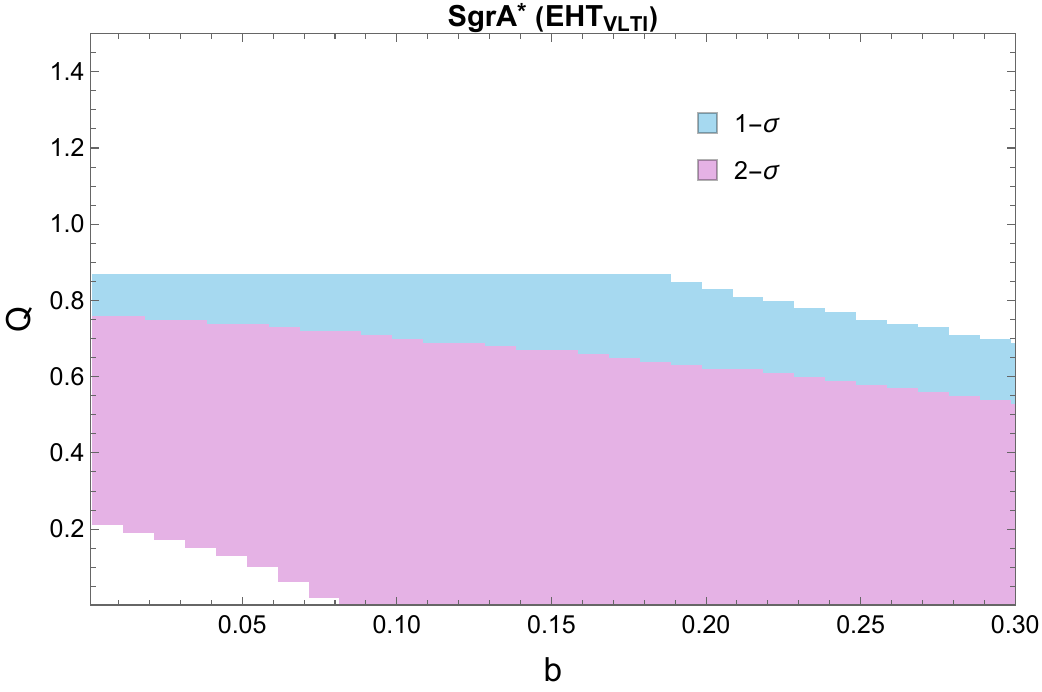}
\includegraphics[scale=0.295]{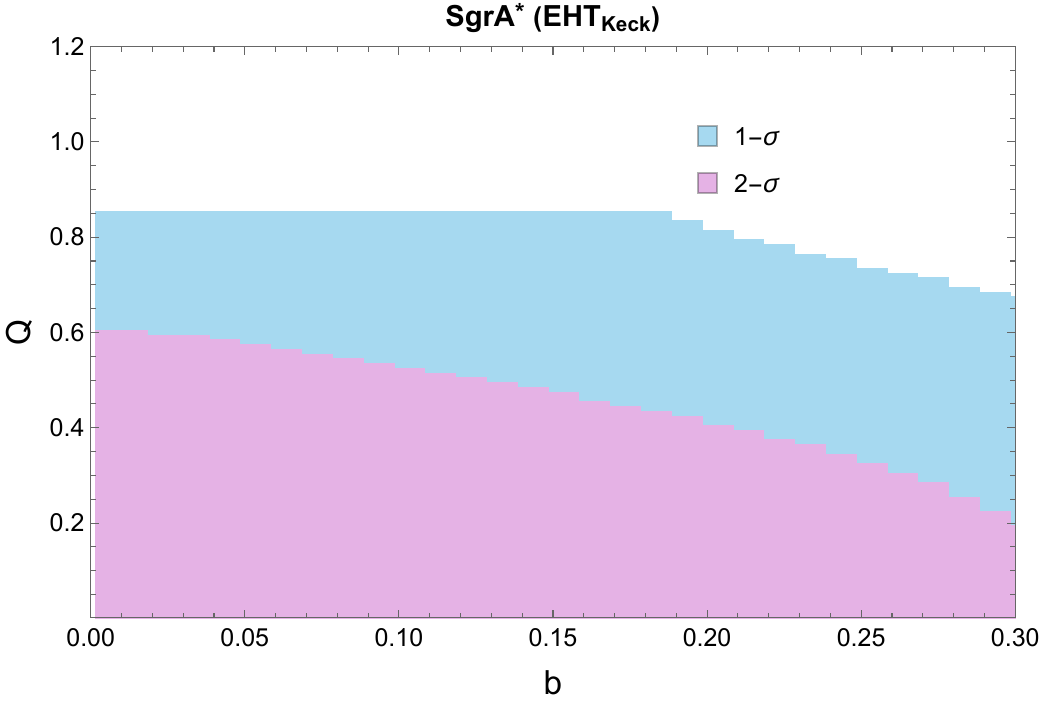}
\vspace{0.2cm}

\includegraphics[scale=0.27]{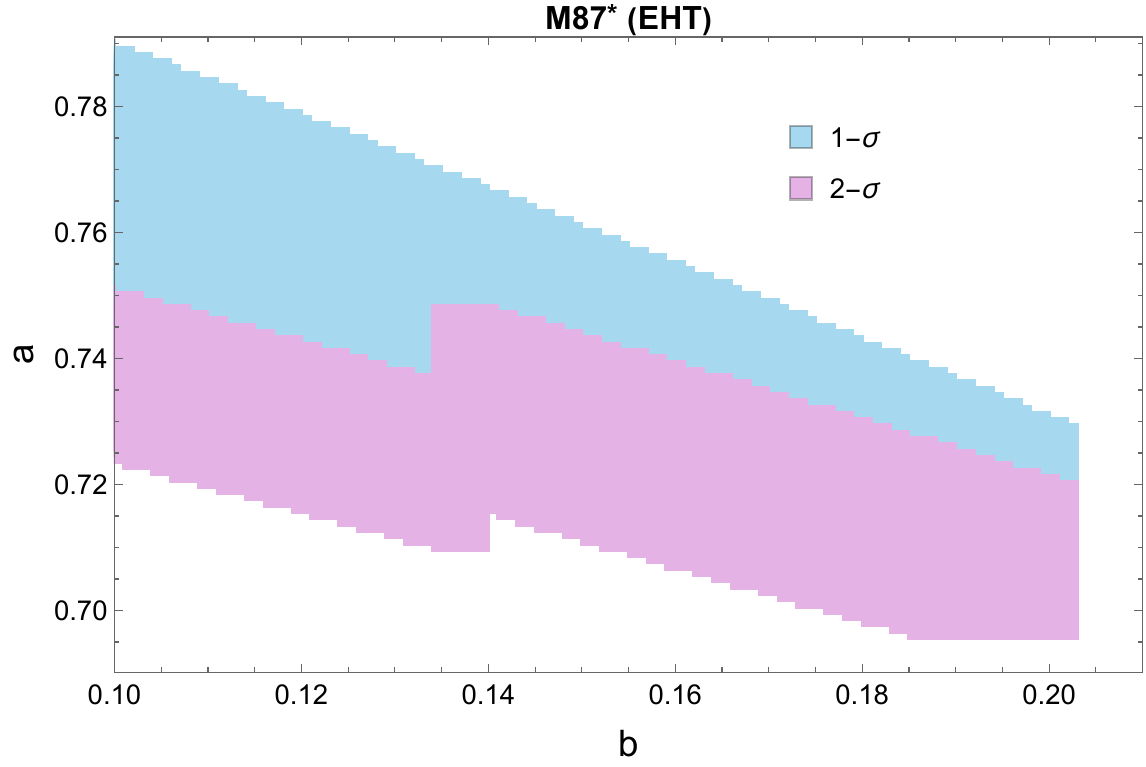}
\includegraphics[scale=0.27]{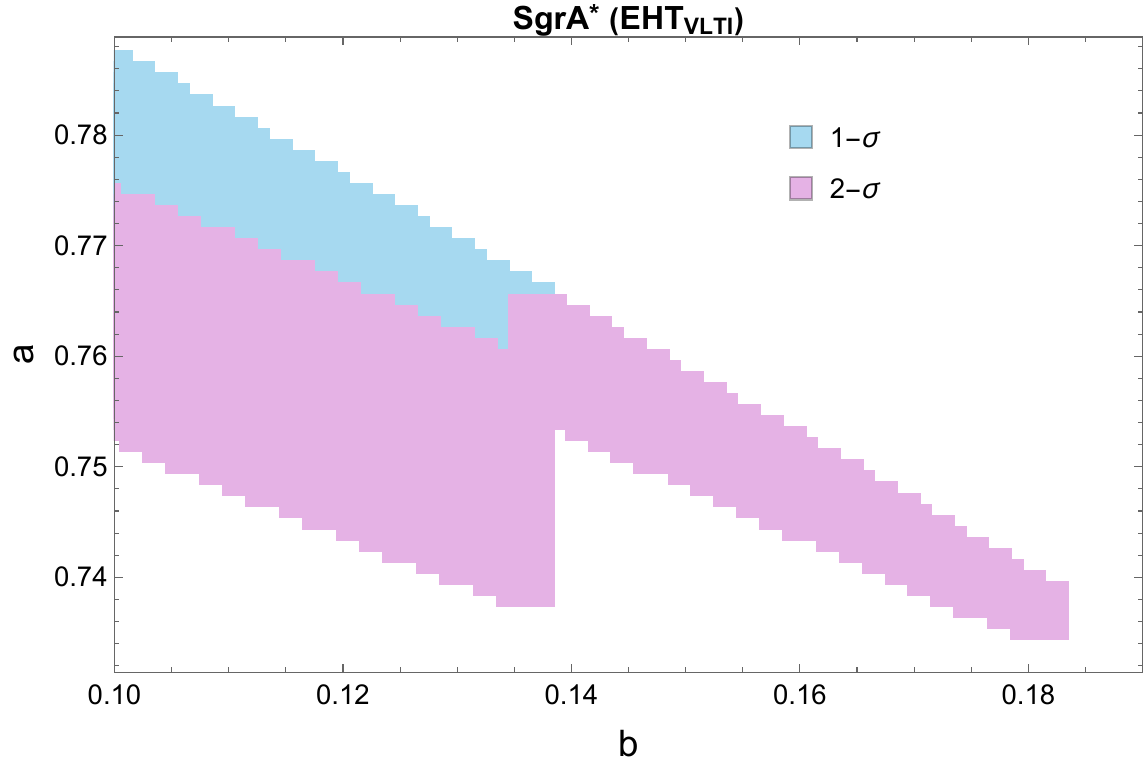}
\includegraphics[scale=0.27]{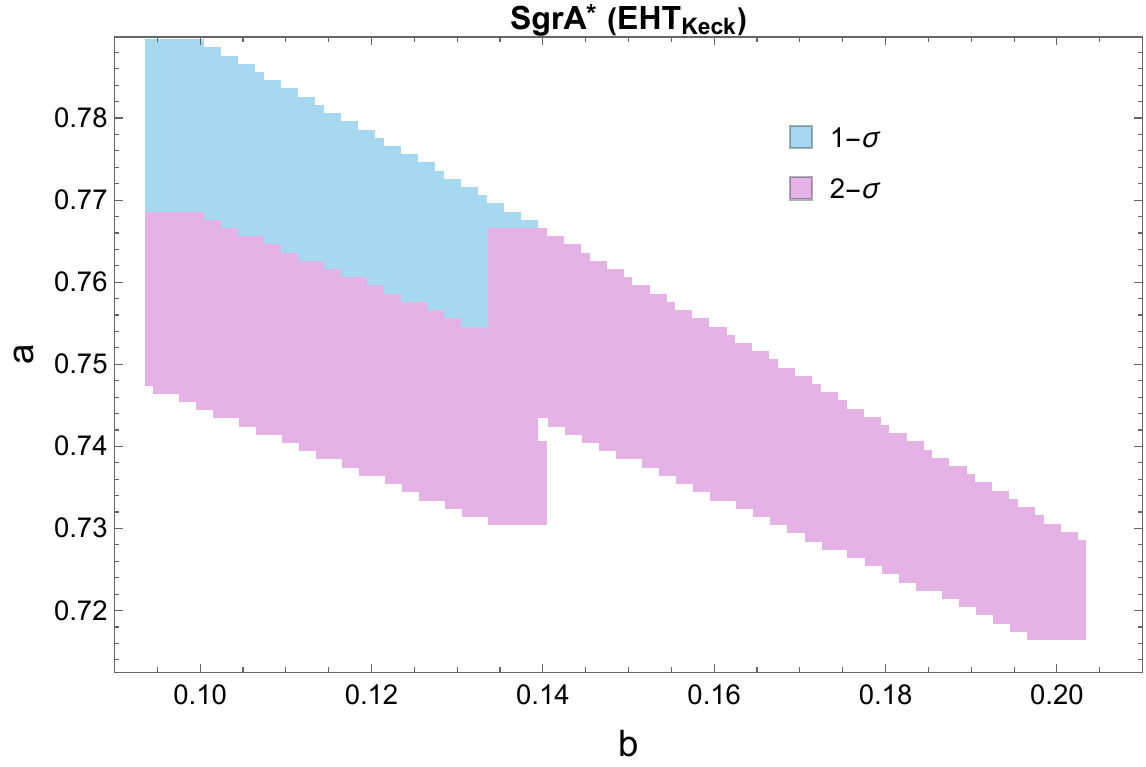}
\caption{\textit{\footnotesize Combined constraint regions in the parameter spaces $(\alpha,b)$, $(Q,b)$,$(a,b)$, and $(N,b)$ obtained from CUDA-based simulations. }}
\label{fig:Moduli}
\end{figure}

In the following discussion, we present a numerical procedure based on CUDA parallel computations to determine the parameter pairs $(\alpha, b)$, $(N, b)$, $(a, b)$, and $(Q, b)$ that yield black hole shadow configurations consistent with  EHT observational data. In each case, all parameters are fixed except for $b$, which is the parameter we aim to constrain. The second parameter in each pair is varied within the allowed range of the model parameter space. For each choice of parameters, the maximal shadow radius $R_{\mathrm{max}}$  has been  computed using a CUDA-based code, enabling an efficient evaluation over a dense grid in the  parameter space. The resulting shadow radius is then compared with the observational bounds provided by EHT  collaboration. This comparison identifies  the parameter space regions  that meet   with the $1-\sigma$ and $2-\sigma$ confidence interval conditions. The  numerical computations  are given  in Fig.~(\ref{fig:Moduli}).

As illustrated in Fig.~(\ref{fig:Moduli}), the regions of the reduced parameter spaces $(\alpha,b)$,  $(a,b)$,  $(Q,b)$, and $(N,b)$ that are consistent with observational data tend to expand for larger values of the parameters. This behavior shows that the black hole  spacetime can reproduce the shadow signatures observed in a relatively large range of configurations, demonstrating the flexibility of the model.  Due to links between parameters, in each analysis one parameter is fixed while the other is varied within its allowed range. In all considered parameter subspaces $(\alpha,b)$,  $(Q,b)$, $(a,b)$, and $(N,b)$, the metric admits wide ranges of  the parameter values  being  compatible with the existence of real and physically meaningful horizons. Confronted with observational data,  however, these ranges become significantly constrained.

The corresponding constraints for M87* and Sgr A* at the $1-\sigma$ and $2-\sigma$ confidence levels are summarized in Table~(\ref{tab11}), together with the results  being obtained for all parameter spaces. In this analysis, we fix $w = -2/3$ throughout, and adopt the reference values $(a = 0.5, Q = 0.5, N = 0.1, \alpha = 0.1)$. For each figure, only one parameter is varied within its allowed range, while the remaining parameters are kept fixed.

\begin{table}[!ht]
\centering
\caption{Constraints on the parameter $b$ for different fixed values of $\alpha$, $a$, $N$, and $Q$, derived from observational bounds at the $1-\sigma$ and $2-\sigma$ confidence levels.}
\label{tab11}
\begin{tabular}{llccc}
\toprule
\textbf{Parameter} & \textbf{Confidence} & \textbf{$M87^*$} & \textbf{$SgrA^*_{\mathrm{VLTI}}$} & \textbf{$SgrA^*_{\mathrm{Keck}}$} \\
\midrule

\multirow{2}{*}{$\alpha = 0.04$}
& $1-\sigma$ & $0.01 < b < 0.44$ & $0.38 < b < 0.44$ & $0.01 < b < 0.44$ \\
& $2-\sigma$ & $0.01 < b < 0.44$ & $0.01 < b < 0.44$ & $0.10 < b < 0.44$ \\

\midrule

\multirow{2}{*}{$N = 0.035$}
& $1-\sigma$ & $0.01 < b < 0.20$ & $0.01 < b < 0.20$ & $0.01 < b < 0.20$ \\
& $2-\sigma$ & $0.01 < b < 0.20$ & $0.14 < b < 0.20$ & $0.19 < b < 0.20$ \\

\midrule

\multirow{2}{*}{$Q = 0.7$}
& $1-\sigma$ & $0.01 < b < 0.28$ & $0.01 < b < 0.30$ & $0.01 < b < 0.25$ \\
& $2-\sigma$ & $0.01 < b < 0.20$ & $0.01 < b < 0.10$ & $0.01 < b < 0.30$ \\
\midrule

\multirow{2}{*}{$a = 0.76$}
& $1-\sigma$ & $0.01 < b < 0.145$ & $0.13 < b < 0.136$ & $0.11 < b < 0.132$ \\
& $2-\sigma$ & $0.01 < b < 0.145$ & $0.1 < b < 0.142$ & $0.1 < b < 0.42$ \\
\bottomrule
\end{tabular}
\end{table}
A clear conclusion emerges from this examination. Indeed, although the parameter space theoretically allows for large regions, only limited regions are compatible with the observational constraints. In particular, the  NC parameter $b$ plays a major role, as it is consistently constrained by the data in all cases, whereas the other parameters primarily influence the shape and  the size of the allowed regions. This indicates that $b$ behaves as a control parameter governing the agreement between the theoretical model and the observed black hole shadows.

\section{Machine learning constraints on  NC black hole parameters from EHT observations}
In this section, we would like  to apply machine learning techniques to the shadow activities.  This approach  has been  motivated by recent developments in the application of machine learning to the physics of  black hole physics and  superstring theory compactifications on Calabi–Yau manifolds.   For instance, these   methods  have been  exploited  in the classical classification of black holes in the mass–spin diagram as well as in quantum gravitational settings \cite{E34,E35}.
In the context of string theory compactifications, machine learning methods have been successfully employed to analyze and classify complex geometric data, including complete intersection Calabi–Yau manifolds and their free quotients using neural networks \cite{E36}.  Moreover,   certain   studies demonstrate that machine learning models are capable of efficiently identifying physical regimes and structural properties of black holes based on their parameter space. In particular, neural networks can detect transition boundaries between different classes of black holes, predict regions of stability, and uncover hidden geometric structures. Supported by such  developments, we employ machine learning techniques, specifically feed-forward FCNN \cite{E37}, to generate black hole shadow configurations consistent with observational data from the EHT.     A priori, there are many models to follows. However, here,   we exploit  a FCNN since it is  a  simple. Moreover, it is a  fast model that can learn the relation between  the physical parameters and  the observational constraints \cite{ML1,ML2}. It has been  well suited for problems with a small number of input parameters giving  stable results with low computational costs. For  such a  reason, it is useful  to probe  large regions of the black  hole  parameter space and  identify configurations matching  with the EHT data. This  can be achieved by learning the underlying relationships between physical parameters and observational constraints.

\subsection{Data preparation}
The datasets used in the present work  are generated through CUDA-based numerical simulations. For each combination of  the parameters $(b, N, \alpha)$, the corresponding black hole shadow  is  approached  and then confronted with observational data provided by  EHT collaboration. This procedure is repeated under various observational constraints, resulting in distinct datasets corresponding to the M87$^*$, Sgr A$^*_{\text{VLTI}}$, and Sgr A$^*_{\text{Keck}}$ cases, each evaluated within the $1\sigma$ and $2\sigma$ confidence intervals. Based on this comparison, each sample is assigned a binary label. A value of 1 is assigned when the calculated shadow is consistent with the EHT  observational limits at the given confidence level, while a value of 0 is given when it is inconsistent.
\begin{equation}
(b,N,\alpha)
\rightarrow \text{1 or 0}.
\end{equation}
In order to ensure the quality of the dataset, a filtering step is applied to remove non-physical or redundant configurations, retaining only parameter values within the physically allowed ranges. The features are then normalized using  a standard scaling to achieve zero mean and unit variance, which improves the convergence and  the stability of the training process. The dataset is subsequently divided into training, validation, and test subsets, ensuring that each black hole configuration appears in only one subset. This helps prevent data leaks and allows models to adapt effectively to unknown configurations.  In fact, this process yields a clean, balanced, and well-structured dataset suitable for classification, making it possible for machine learning models to accurately determine whether a given set of parameters produces a black hole shadow consistent with the EHT  observational data.

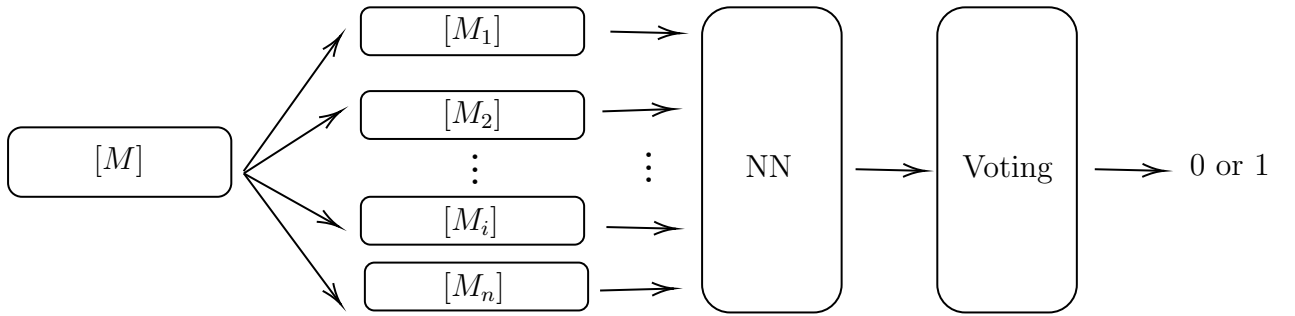
\begin{figure}[h!]
\centering
\begin{tikzpicture}[x=0.75pt,y=0.75pt,yscale=-1,xscale=1]

\draw (131.09,122.09) -- (178.63,186.21);
\draw [shift={(179.82,187.82)}, rotate=233.45, line width=0.75]
      (10.93,-3.29) .. controls (6.95,-1.4) and (3.31,-0.3) .. (0,0)
      .. controls (3.31,0.3) and (6.95,1.4) .. (10.93,3.29);

\draw (131.09,122.09) -- (177.07,147.84);
\draw [shift={(178.82,148.82)}, rotate=209.25, line width=0.75]
      (10.93,-3.29) .. controls (6.95,-1.4) and (3.31,-0.3) .. (0,0)
      .. controls (3.31,0.3) and (6.95,1.4) .. (10.93,3.29);

\draw (131.09,122.09) -- (178.14,91.9);
\draw [shift={(179.82,90.82)}, rotate=147.31, line width=0.75]
      (10.93,-3.29) .. controls (6.95,-1.4) and (3.31,-0.3) .. (0,0)
      .. controls (3.31,0.3) and (6.95,1.4) .. (10.93,3.29);

\draw (131,121) -- (178.65,54.44);
\draw [shift={(179.82,52.82)}, rotate=125.6, line width=0.75]
      (10.93,-3.29) .. controls (6.95,-1.4) and (3.31,-0.3) .. (0,0)
      .. controls (3.31,0.3) and (6.95,1.4) .. (10.93,3.29);

\draw (315,51) -- (347.82,51.77);
\draw [shift={(349.82,51.82)}, rotate=181.35, line width=0.75]
      (10.93,-3.29) .. controls (6.95,-1.4) and (3.31,-0.3) .. (0,0)
      .. controls (3.31,0.3) and (6.95,1.4) .. (10.93,3.29);

\draw (311,91) -- (344.82,89.88);
\draw [shift={(346.82,89.82)}, rotate=178.11, line width=0.75]
      (10.93,-3.29) .. controls (6.95,-1.4) and (3.31,-0.3) .. (0,0)
      .. controls (3.31,0.3) and (6.95,1.4) .. (10.93,3.29);

\draw (309.82,180.82) -- (344.82,179.87);
\draw [shift={(346.82,179.82)}, rotate=178.45, line width=0.75]
      (10.93,-3.29) .. controls (6.95,-1.4) and (3.31,-0.3) .. (0,0)
      .. controls (3.31,0.3) and (6.95,1.4) .. (10.93,3.29);

\draw (313,149) -- (345.82,149.77);
\draw [shift={(347.82,149.82)}, rotate=181.35, line width=0.75]
      (10.93,-3.29) .. controls (6.95,-1.4) and (3.31,-0.3) .. (0,0)
      .. controls (3.31,0.3) and (6.95,1.4) .. (10.93,3.29);

\draw (438,120) -- (470.82,120.77);
\draw [shift={(472.82,120.82)}, rotate=181.35, line width=0.75]
      (10.93,-3.29) .. controls (6.95,-1.4) and (3.31,-0.3) .. (0,0)
      .. controls (3.31,0.3) and (6.95,1.4) .. (10.93,3.29);

\draw (558,120) -- (590.82,120.77);
\draw [shift={(592.82,120.82)}, rotate=181.35, line width=0.75]
      (10.93,-3.29) .. controls (6.95,-1.4) and (3.31,-0.3) .. (0,0)
      .. controls (3.31,0.3) and (6.95,1.4) .. (10.93,3.29);

\draw (12.82,105.96) .. controls (12.82,102.12) and (15.94,99) .. (19.78,99) -- (117.85,99)
      .. controls (121.7,99) and (124.82,102.12) .. (124.82,105.96) -- (124.82,126.85)
      .. controls (124.82,130.7) and (121.7,133.82) .. (117.85,133.82) -- (19.78,133.82)
      .. controls (15.94,133.82) and (12.82,130.7) .. (12.82,126.85) -- cycle;

\draw (189.82,43.62) .. controls (189.82,40.97) and (191.97,38.82) .. (194.62,38.82)
      -- (297.02,38.82) .. controls (299.67,38.82) and (301.82,40.97) .. (301.82,43.62)
      -- (301.82,58.02) .. controls (301.82,60.67) and (299.67,62.82) .. (297.02,62.82)
      -- (194.62,62.82) .. controls (191.97,62.82) and (189.82,60.67) .. (189.82,58.02) -- cycle;

\draw (191.82,171.62) .. controls (191.82,168.97) and (193.97,166.82) .. (196.62,166.82)
      -- (299.02,166.82) .. controls (301.67,166.82) and (303.82,168.97) .. (303.82,171.62)
      -- (303.82,186.02) .. controls (303.82,188.67) and (301.67,190.82) .. (299.02,190.82)
      -- (196.62,190.82) .. controls (193.97,190.82) and (191.82,188.67) .. (191.82,186.02) -- cycle;

\draw (189.82,85.62) .. controls (189.82,82.97) and (191.97,80.82) .. (194.62,80.82)
      -- (297.02,80.82) .. controls (299.67,80.82) and (301.82,82.97) .. (301.82,85.62)
      -- (301.82,100.02) .. controls (301.82,102.67) and (299.67,104.82) .. (297.02,104.82)
      -- (194.62,104.82) .. controls (191.97,104.82) and (189.82,102.67) .. (189.82,100.02) -- cycle;

\draw (189.82,138.62) .. controls (189.82,135.97) and (191.97,133.82) .. (194.62,133.82)
      -- (297.02,133.82) .. controls (299.67,133.82) and (301.82,135.97) .. (301.82,138.62)
      -- (301.82,153.02) .. controls (301.82,155.67) and (299.67,157.82) .. (297.02,157.82)
      -- (194.62,157.82) .. controls (191.97,157.82) and (189.82,155.67) .. (189.82,153.02) -- cycle;

\draw (361,52.82) .. controls (361,45.09) and (367.27,38.82) .. (375,38.82)
      -- (417,38.82) .. controls (424.73,38.82) and (431,45.09) .. (431,52.82)
      -- (431,177.82) .. controls (431,185.55) and (424.73,191.82) .. (417,191.82)
      -- (375,191.82) .. controls (367.27,191.82) and (361,185.55) .. (361,177.82) -- cycle;

\draw (479,52.82) .. controls (479,45.09) and (485.27,38.82) .. (493,38.82)
      -- (535,38.82) .. controls (542.73,38.82) and (549,45.09) .. (549,52.82)
      -- (549,177.82) .. controls (549,185.55) and (542.73,191.82) .. (535,191.82)
      -- (493,191.82) .. controls (485.27,191.82) and (479,185.55) .. (479,177.82) -- cycle;

\draw (50,102) node [anchor=north west] { $\displaystyle [M]$};
\draw (225,37) node [anchor=north west] {$\displaystyle \left[M_1 \right]$};
\draw (225,79) node [anchor=north west] {$\displaystyle \left[M_2 \right]$};
\draw (225,132.82) node [anchor=north west] {$\displaystyle \left[M_i \right]$};
\draw (225,166) node [anchor=north west] {$\displaystyle \left[M_n \right]$};

\draw (239,107) node [anchor=north west] {\shortstack{\textbf{.}\\\textbf{.}\\\textbf{.}}};
\draw (326,106) node [anchor=north west] {\shortstack{\textbf{.}\\\textbf{.}\\\textbf{.}}};

\draw (377,107) node [anchor=north west] {NN};
\draw (486,107) node [anchor=north west] {Voting};
\draw (600,107) node [anchor=north west] {0 or 1};

\end{tikzpicture}
\caption{Voting procedure for consistency of the shadow with EHT observations} \label{77}
\end{figure}

To train the machine learning models, we employ a feedforward neural NN that takes the input tuples and predicts the corresponding binary label (0 or 1). To improve accuracy and reduce sensitivity to small variations in the input parameters, we adopt a voting strategy. In this approach, for each input tuple $[M]=(b,N,\alpha)$, several equivalent representations $[M_1,M_2,\ldots,M_n]$  can be  generated. Each representation is processed independently by the NN. Moreover, the final prediction  can be  determined by a majority vote over the outputs.  More precisely, the voting procedure generates several small variations of the same physical input while keeping the main physical properties unchanged. Each version  has been sent independently to the FCNN model, which  generates a binary prediction. The final result is then considered  as the most common output among all predictions. This method provides a  classification  being more stable and less sensitive to small numerical changes. This can improve   the reliability of the final decision. This procedure, illustrated in Fig.~(\ref{77}), ensures  a robust classification. Moreover, it  improves a generalization  via  different black hole configurations.

As shown in Fig.~(\ref{77}), each input parameter set is first processed by the neural network, which generates predictions for the corresponding configurations. These predictions are then combined using a voting scheme to produce the final output label, indicating whether the resulting black hole shadow is consistent with EHT observations. This combination of careful data preparation and an advanced learning algorithm yields a robust and accurate classification framework.

\subsection{FCNN classification of  NC black hole shadow consistency with EHT observations} 
In this part, we examine the consistency of the  black hole shadow configurations with observational data from the EHT using a FCNN. The dataset consists of $20{,}000$ samples, where each sample is characterized by the set of  physical parameters  $
(b, N, \alpha).$
To ensure a balanced dataset, we  should select an equal number of samples being  consistent and inconsistent with EHT observational constraints. After shuffling, the data are divided into
\[
70\% \text{ training}, \qquad 20\% \text{ validation}, \qquad 10\% \text{ testing}.
\]
To classify whether a given parameter configuration produces a black hole shadow compatible with EHT observations, we employ a  FCNN model  with a two-class probabilistic output. The network architecture is schemed  as follows 
\begin{equation}
N_{3 \times 2} = \bigl(F \circ G_{3 \times 32},\; F \circ G_{32 \times 64},\; F \circ G_{64 \times 32},\; S \circ G_{32 \times 2}\bigr),
\end{equation}
where $G_{n \times m}$ denotes a fully connected layer with $n$ inputs and $m$ outputs. $F$ represents the ReLU activation function, and $S$ is a softmax layer producing the final class probabilities.

\begin{figure}[!h]
\centering
\includegraphics[scale=0.35]{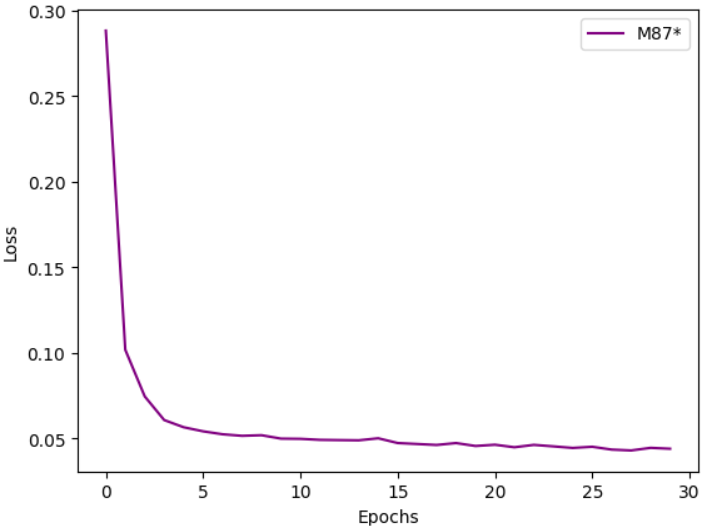}\hspace{2mm}
\includegraphics[scale=0.35]{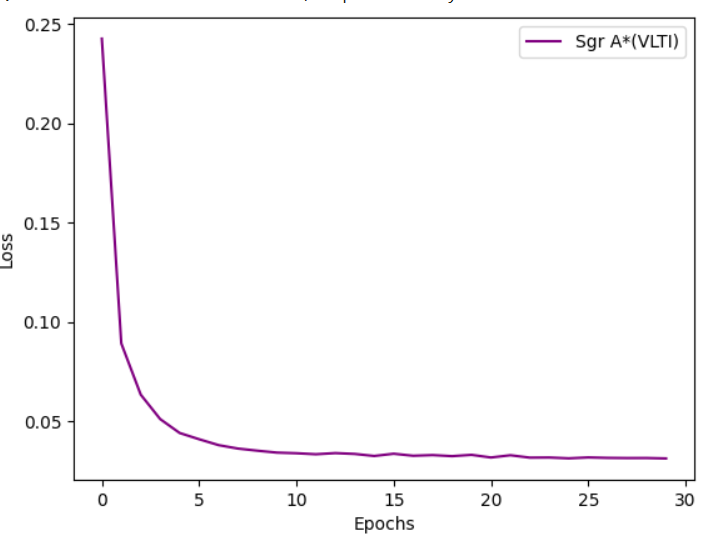}\hspace{2mm}
\includegraphics[scale=0.35]{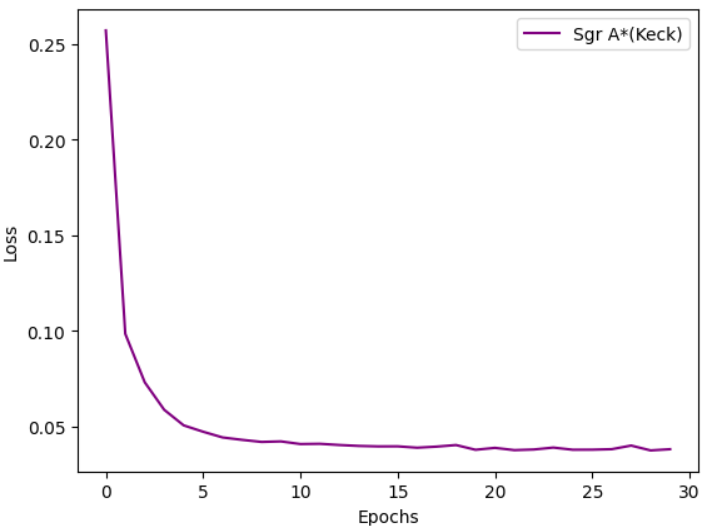}
\caption{Training curves of the FCNN model for the $1-\sigma$ observational constraint.}
\label{fig:train}
\end{figure}

\begin{figure}[!h]
\centering
\includegraphics[scale=0.35]{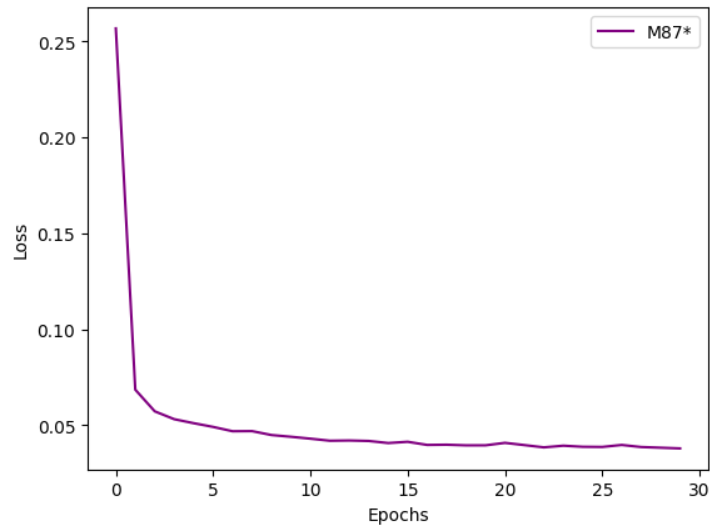}\hspace{1mm}
\includegraphics[scale=0.35]{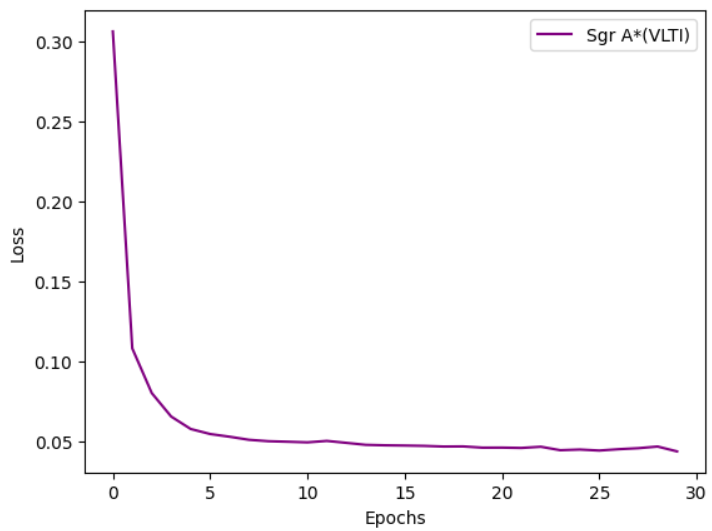}\hspace{1mm}
\includegraphics[scale=0.35]{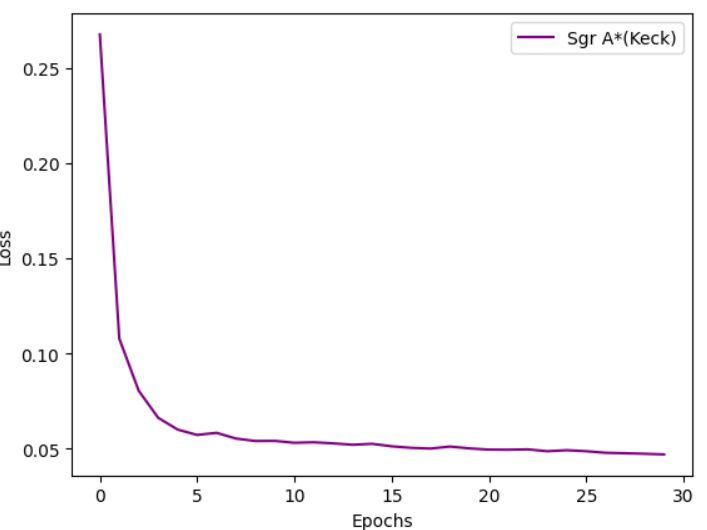}
\caption{Training curves of the FCNN model for the $2-\sigma$ observational constraint.}
\label{fig:train1}
\end{figure}

The model is trained using the Adam optimizer with a binary cross-entropy loss function. We train the neural network using $2,000$ data samples. During training, the learning rate is automatically adjusted to ensure the  stability and the efficient convergence. Such behaviors are illustrated   in Figs.~(\ref{fig:train}) and (\ref{fig:train1}) for    $1-\sigma$  and  $2-\sigma$ observational constraints, respectively. 

As shown from these figures, the loss decreases rapidly during the initial epochs and then stabilizes. This  indicates efficient learning and early convergence behaviors. The training and the  validation accuracies remain consistently high across all datasets. This  reveals   a good generalization performance without signs of overfitting. The small fluctuations  being observed in the validation loss  have been  expected and arise from a  stochastic mini-batch optimization. However, they do not affect the global  convergence behaviors. The model is further evaluated on independent test sets. Moreover,  the accuracy  has been  computed separately for each observational case. The  obtained results confirm the robustness of the classifier across all configurations.  In fact,  the network successfully distinguishes between configurations  being  compatible and incompatible with EHT observational constraints. This shows   that it  has learned a well-defined decision boundary in the three-dimensional parameter space coordinated by the NC black hole moduli.  Roughly,  the  detailed performance metrics are illustrated  in Tables~(\ref{tab1}) and (\ref{tab2}) for   $1-\sigma$  and  $2-\sigma$ observational constraints, respectively.

\begin{table}[!ht]
\centering
\caption{Model performance of the FCNN model  for the $1-\sigma$ observational constraint.}
\label{tab1}
\begin{tabular}{lccc}
\toprule
\textbf{Metric} & $M87^*$ & $SgrA^*_{\mathrm{VLTI}}$ & $SgrA^*_{\mathrm{Keck}}$ \\
\midrule
Test Accuracy & 0.978 &  0.9779 & 0.981 \\
Voting Accuracy & 0.979 & 0.98 & 0.9815 \\
\bottomrule
\end{tabular}
\end{table}

\begin{table}[!ht]
\centering
\caption{Model performance of the FCNN model  for the $2-\sigma$ observational constraint.}
\label{tab2}
\begin{tabular}{lccc}
\toprule
\textbf{Metric} & $M87^*$ & $SgrA^*_{\mathrm{VLTI}}$ & $SgrA^*_{\mathrm{Keck}}$ \\
\midrule
Test Accuracy &  0.9804 & 0.9794 & 0.9769 \\
Voting Accuracy & 0.9815 & 0.9795 & 0.9772 \\
\bottomrule
\end{tabular}
\end{table}

To evaluate the  robustness, we employ a voting system inspired by symmetry-based classification methods. For each configuration, we generate 100 perturbed samples while preserving their symmetry, and then  we classify them independently. The final prediction can be obtained by majority voting, allowing  to test the model stability in the face of structured perturbations. This method provides high and consistent voting accuracy across all datasets. This confirms that the classifier remains stable in the face of small deformations that preserve  the symmetry. Furthermore, it generalizes the training distribution.

 Although global metrics such as test accuracy and voting accuracy provide a global evaluation of the model performance, they do not offer detailed  data  about the distribution of classification results.  To evince these issues, we approach the confusion matrices.  Precisely, they are illustrated in Tables~(\ref{tab_cm1}) and (\ref{tab_cm2}) for   $1-\sigma$  and  $2-\sigma$ observational constraints, respectively.  Indeed, they allow a more detailed analysis of  how the model assigns predictions across the two classes by explicitly showing correct classifications and misclassifications. In particular, these matrices make it possible  not only to quantify the number of errors and  but also to identify any imbalance in the predictions. The results clearly reveals  that the model correctly classifies nearly all samples, with only a negligible number of misclassifications.  Thus, the confusion matrices provide a more complete and transparent assessment of the classifier performance beyond aggregate metrics.

\begin{table}[ht!]
\centering
\caption{Confusion matrix using the FCNN model for the $1-\sigma$ observational constraint.}
\label{tab_cm1}
\begin{tabular}{|c|cc|cc|cc|}
\hline
 & \multicolumn{2}{c|}{$M87^*$} 
 & \multicolumn{2}{c|}{$SgrA^*_{\mathrm{VLTI}}$} 
 & \multicolumn{2}{c|}{$SgrA^*_{\mathrm{Keck}}$}  \\ 
\hline
 & Class 0 & Class 1 & Class 0 & Class 1 & Class 0 & Class 1 \\ 
\hline
Class 0 
 & 1,425 & 15 & 1,854 & 13 & 1,470 & 24 \\ 
Class 1 
 & 27 & 533 & 24 & 109 & 16 & 490 \\ 
\hline
\end{tabular}
\end{table}

\begin{table}[ht!]
\centering
\caption{Confusion matrix using the  FCNN model for the $2-\sigma$ observational constraint.}
\label{tab_cm2}
\begin{tabular}{|c|cc|cc|cc|}
\hline
 & \multicolumn{2}{c|}{$M87^*$} 
 & \multicolumn{2}{c|}{$SgrA^*_{\mathrm{VLTI}}$} 
 & \multicolumn{2}{c|}{$SgrA^*_{\mathrm{Keck}}$}  \\ 
\hline
 & Class 0 & Class 1 & Class 0 & Class 1 & Class 0 & Class 1 \\ 
\hline
Class 0 
 & 996 & 27 & 1,597 & 19 & 1,736 & 27 \\ 
Class 1 
 & 10 & 967 & 22 & 362 & 14 & 223 \\ 
\hline
\end{tabular}
\end{table}
A close examination  reveals that the $1-\sigma$ observational constraint  provides slightly better and more stable performance than the $2-\sigma$ case, particularly for the $SgrA^*_{\mathrm{Keck}}$ dataset, where the highest accuracies are consistently obtained.  Concerning  the $2-\sigma$ constraint, however, a more extensive uncertainty range  has been   implemented. This  leads  to a slight decrease in  the classification performance,  being  noticeable in the $SgrA^*_{\mathrm{Keck}}$  empirical findings. Thus, the model is more strictly constrained and better fitted under the $1-\sigma$ observation regime, while remaining robust in the more flexible  $2-\sigma$ scenario.
\section{Concluding remarks}

In this paper, we have investigated   the shadow  behaviors  of NC rotating and charged  black holes with dark energy sectors and a cloud of strings by combining machine learning techniques, EHT observations, and CUDA-based numerical methods. First, we have analyzed the horizon structure through the metric functions of such black hole solutions, which encode  the relevant shadow properties of the solutions. Then, we have applied the Hamilton--Jacobi formalism together with CUDA-accelerated simulations to determine the one-dimensional shadow curves and the energy emission rates in terms of  black hole parameter variations. More specifically, we have  revealed  that the  rotating parameter $a$ and  the charge $Q$ affect    both the size of the shadow and the energy emission rate.  For large values of the parameter $N$ and small values of $\alpha$, we have  found that the shadow has  developed   a D-like shape,  showing  the impact of the NC parameter $b$, which controls both the size and the global shape of the NC black hole shadows. After that,  we have  elaborated  a CUDA-based numerical framework to constrain the NC  black hole parameters through a direct comparison with astrophysical observations, including those reported by the EHT collaboration. Using this framework, we have shown that the NC parameter $b$ remains positive and below approximately $0.44$ in order to be consistent with these observations. The obtained results  have been  then implemented within machine learning techniques to generate reliable training data for a FCNN  exploited  to  bridge the  obtained  theoretical predictions with the shadow observations reported by the EHT collaboration. As a result, we have demonstrated that the fully connected and trained neural network accurately identified the shadow observations reported by the EHT collaboration. A global comparison has  revealed that the $1-\sigma$ observational constraint  has yielded slightly better and more stable performance than the $2-\sigma$ case, particularly for the $SgrA^*_{\mathrm{Keck}}$ dataset, where the highest accuracies have been  consistently obtained. Under the $2-\sigma$ constraint,  we have observed that the larger uncertainty range slightly reduced the classification performance, especially for the $SgrA^*_{\mathrm{Keck}}$ dataset. These findings  have provided   the potential of CUDA-accelerated machine learning for constraining  black hole parameters  in non-trivial geometries including NC ones.  Although the present machine learning approach supplies  strong classification results, it should be interpreted in a cautious way. In future works, it would be useful to test more advanced learning methods. Moreover, it could be possible to add standard evaluation tools such as ROC curves and precision-recall analysis to better check the performance and stability of the  proposed model. It would also be important to provide   more details  concerning  the voting procedure exploited  in the classification process. 

This work raises several open questions for future investigations. In particular, it would be interesting to further refine the analysis of the limitations linked to the exclusive use of the shadow radius compared to EHT data, as well as to examine more closely the role of observational uncertainties and theoretical modeling assumptions in the reliability of the results.

{\bf Data availability}\\
  Data sharing is not applicable to this article. \\
  
{\bf Funding information}\\
  It is  not applicable.

\section*{Acknowledgements}

The author would like to thank the Department of Mathematics at the University of Aveiro and the Center for Research and Development
in Mathematics and Applications (CIDMA) in Aveiro for their kind hospitality during the preparation of this work.  She is very
grateful to  Carlos A. R. Herdeiro  for  his scientific support, discussions, suggestions,   and encouragement.  She would like to thank J. Ferreira, E. C. Filho, M. Mariano,  J. Nicoules,  H. Olivares,     G. Ribeiro, J. J. Vázquez, and M. Wielgus   for their discussions and enjoyable company. 
 She thanks A. Belhaj for the discussions, encouragement, supervising,  and collaboration on related topics. She would also like to thank S. E. Baddis, H. Belmahi, and S. E. Ennadifi for their discussions and collaborations.
The author is very grateful to the CNRST for its financial support under
the PASS doctoral fellowship program.


\begin{thebibliography}{99}
\bibitem{ii4}
M. Jemri, \textit{Dunkl-Corrected Deformation of RN-AdS Black Hole Thermodynamics,} Theoretical and Mathematical Physics 227, 1 (2026), \texttt{	arXiv:2512.19200 [hep-th]}.
\bibitem{i1} H. Belmahi, M. Jemri, R. Salih, \textit{Stability and Criticality Behaviors of Accelerating Charged AdS Black Holes in Rainbow Gravity,} International Journal of Modern Physics A 40, 35 (2025), \texttt{arXiv:2507.03572 [gr-qc]}.

\bibitem{i2} S. E. Baddis, A. Belhaj, H. Belmahi, W. El Hadri, M. Jemri, \textit{On RN-AdS Black Holes with a Cloud of Strings and Quintessence in Noncommuative Geometry,}   \texttt{arXiv:2509.11356 [hep-th]}.


\bibitem{i3}
T. Clifton, P. G. Ferreira, A. Padilla, C. Skordis, \textit{Modified Gravity and Cosmology,} Phys.Rept. 513, 1 (2012), \texttt{	arXiv:1106.2476 [astro-ph.CO].}

\bibitem{i3333}
C. Herdeiro, E. Radu , E. d. S. C. Filho , N. Sanchis-Gual, \textit{Multipolar Proca stars:
electric, magnetic and hybrid solitons,}\texttt{		arXiv:2605.13965 [gr-qc].}



\bibitem{ii5}
A. Belhaj, M. Jemri, \textit{On Thermodynamics of Charged Black Holes via Extended Space-time Derivatives,} International Journal of Modern Physics A 41, 04 (2026), \texttt{		arXiv:2511.18407 [hep-th].}



\bibitem{i4}
P. Nicolini, A. Smailagic, E. Spallucci \textit{Noncommutative geometry inspired Schwarzschild black hole,} Phys.Lett.B 632, 547 (2006), \texttt{			arXiv:gr-qc/0510112.}

\bibitem{i5}
W. El Hadri, M. Jemri, \textit{Thermodynamics and Criticality of Noncommutative RN-AdS Black Holes,}  Braz J Phys 56 23 (2026), \texttt{	arXiv:2509.00926 [hep-th].}

\bibitem{i6}
S. E. Baddis, A. Belhaj, H. Belmahi, W. El Hadri, M. Jemri, \textit{On RN-AdS Black Holes with a Cloud of Strings and Quintessence in Noncommuative Geometry,}   \texttt{		arXiv:2509.11356 [hep-th].}


\bibitem{i7}
M. A. A. de Paula, H. C. D. Lima, P. V. P. Cunha,  C. A. R. Herdeiro,
 L. C. B. Crispino, \textit{The two shadows of a single black hole:
Vacuum birefringence phenomena within Einstein-Nonlinear-Electrodynamics,}   \texttt{		arXiv:2603.17007 [gr-qc].}

\bibitem{i8}
G. Bertone, \textit{Dark matter, black holes, and gravitational waves,} Nucl.Phys.B 1003, 116487 (2024), \texttt{		arXiv:2404.11513v1 [astro-ph.CO].}

\bibitem{i9}
A. Belhaj, H. Belmahi, M. Benali, H. El Moumni, M. Amine Essebani, M. B. Sedra, \textit{Optical Shadows of Rotating Bardeen AdS Black Holes,} Mod.Phys.Lett.A 37,  2250032 (2022), \texttt{			arXiv:2202.10892 [gr-qc].}

\bibitem{i10}
 A. Belhaj, A. El Balali, W. El Hadri, H. El Moumni, M. B. Sedra, \textit{Dark Energy Effects on Charged and Rotating Black Holes,}  Eur.Phys.J.Plus 134.  9, 422 (2019), \texttt{			arXiv:1912.08687 [hep-th].}
 
 
 \bibitem{10i} R. F. Rosato, S. Biswas, S. Chakraborty, P. Pani, \textit{Excitation factors for horizonless compact objects: long-lived modes, echoes, and greybody factors,} Phys.Rev.D 113  084002 (2026), \texttt{arXiv:2511.08692 [gr-qc].}


\bibitem{11i} R. F. Rosato, S. Biswas, S. Chakraborty, P. Pani, \textit{Greybody factors, reflectionless scattering modes, and echoes of ultracompact horizonless objects,} Phys.Rev.D 111,   084051  (2025), \texttt{arXiv:2501.16433 [gr-qc].}
 
 \bibitem{12i} I. Banerjee, S. Chakraborty, S. SenGupta, \textit{Silhouette of M87*: A new window to peek into the world of hidden dimensions,} Phys. Rev. D 101, 041301 (2020), \texttt{	arXiv:1909.09385 [gr-qc].}
 
 \bibitem{hajar} H. Belmahi, \textit{Constrained Deflection Angle and Shadows of Rotating Black Holes in Einstein-Maxwell-scalar Theory,} International Journal of Geometric Methods in Modern Physics 23, 09, 2550248 (2026), \texttt{	arXiv:2411.11622 [hep-th].} 
 
 
 
  \bibitem{adil} 
 A. Belhaj, M. Benali, A. El Balali, H. El Moumni and S-E. Ennadifi,  \textit{Deflection an-gle and shadow behaviors of quintessential black holes in arbitrary dimensions}, Class. Quantum Grav. 37 (2020) 215004,   {\tt arXiv:2006.01078}.
  \bibitem{carlos} 
 S. V. M. C. B. Xavier, P. V. P. Cunha, L. C. B. Crispino, C. A. R. Herdeiro, \textit{Shadows of charged rotating black holes: Kerr–Newman versus Kerr–Sen}, Int. J. Mod. Phys. D 29 (2020) 2041005,   {\tt arXiv:2003.14349}.
 \bibitem{belhaj}
A.~Belhaj, M.~Benali and Y.~Hassouni, {\it Superentropic black hole shadows in arbitrary dimensions}, Eur. Phys. J. C \textbf{82} (2022) 619, {\tt arXiv:2203.06774}.
 
 
\bibitem{ST1}
A. Belhaj, H. Belmahi, A. Bouhouch, S. E. Ennadifi, M. B. Sedra,
\textit{Black holes and black strings in M-theory on Calabi–Yau threefolds with four Kähler parameters,}
Eur. Phys. J. C 85, 901 (2025),\texttt{	arXiv:2501.07167 [hep-th]}.

\bibitem{ST2}
A. Belhaj, H. Belmahi, M. Benali, W. El Hadri, H. El Moumni, E. Torrente-Lujan,
\textit{Shadows of 5D Black Holes from String Theory,}
Phys. Lett. B 812, 136025 (2020), \texttt{	arXiv:2008.13478 [hep-th]}.

\bibitem{ST3}
A. Belhaj, M. Benali, A. El Balali, W. El Hadri, H. El Moumni, E. Torrente-Lujan,
\textit{Black hole shadows in M-theory scenarios,}
Int. J. Mod. Phys. D 30, 2150026 (2021), \texttt{	arXiv:2008.09908 [hep-th]}.
 
 \bibitem{Q22} H. K. Sudhanshu, D. V. Singh, S. Upadhyay, Deepika, Y. Myrzakulov, and K. Myrzakulov, \textit{Thermodynamic phase structure and shadow analysis of 4D AdS skyrmion black holes,} International Journal of Modern Physics A 41,  14, 2650089 (2026).
 

 \bibitem{ii11}
 K. Akiyama and al., \textit{First M87 Event Horizon Telescope Results. IV. Imaging the Central Supermassive Black Hole}, Astrophys. J. 875, L4 (2019).

 \bibitem{ii12}
K. Akiyama and al., \textit{First M87 Event Horizon Telescope Results. V. Imaging the Central
Supermassive Black Hole,} Astrophys. J. 875 L5 (2019).
 
 \bibitem{ii13}
K. Akiyama and al.,\textit{ First M87 Event Horizon Telescope Results. VI. Imaging the Central Supermassive Black Hole}, Astrophys. J. 875 L6 (2019).



 \bibitem{iii11}
S. E. Baddis, A. Belhaj, H. Belmahi, S. E. Ennadifi, M. Jemri, \textit{On Computational CUDA Studies of Black Hole Shadows},  \texttt{arXiv:2604.14213 }.

 \bibitem{iii12}
S. E. Baddis, A. Belhaj, H. Belmahi, M. Jemri, \textit{Constraining Black Hole Shadows in Dunkl Spacetime using CUDA Numerical Computations,}   
Journal of High Energy Astrophysics, 51,  100541 (2026), \texttt{arXiv:2510.16460 [gr-qc]}.

 \bibitem{o1}
A. Elafrou, G. Thomas Collignon,\textit{ Introduction to CUDA Performance Optimization,} Nvidia.

 \bibitem{o2} Nvidia, \textit{CUDA C++ Programming Guide.}


 \bibitem{o3} S. E. Baddis, A. Belhaj, and H. Belmahi, \textit{CUDA Assisted Swampland and Black Hole
Thermodynamics}, \texttt{arXiv:2508.12378 [hep-th].}
 
  \bibitem{o4} R. Ginjupalli and G. Khanna, \textit{High-Precision Numerical Simulations of Rotating Black
Holes Accelerated by CUDA,} \texttt{arXiv:1006.0663 [physics.comp-ph].}
  
  



\bibitem{i12} J. v. d. Gucht, J. Davelaar, L. Hendriks, O. Porth, H. Olivares, Y. Mizuno, C. M. Fromm, H. Falcke, \textit{Deep Horizon; a machine learning network that recovers accreting black hole parameters} Astron.Astrophys. 636, A94 (2020).

\bibitem{i13} Medeiros et al., \textit{Machine Learning Delivers Sharper Black Hole Image}, Physics 16, 63
(2023).

\bibitem{i14} N. Steinle and S. Safi-Harb,\textit{ Machine learning classification of black holes in the
mass–spin diagram}, Phys. Rev. D 112, 103038 (2025).




\bibitem{100} P. Nicolini, A. Smailagic, E. Spallucci, \textit{Noncommutative geometry inspired Schwarzschild
black hole.} Physics Letters B, 632 547 (2006).


\bibitem{101} M. R. Douglas, C. M. Hull, \textit{D-branes and the noncommutative torus,} Journal of High
Energy Physics, 02 008 (1998).

\bibitem{NC1}
A. Belhaj, M. Hssaini, E. L. Sahraoui, E. H. Saidi, \textit{Explicit Derivation of Yang-Mills Self-Dual Solutions on non-Commutative Harmonic Space},  Class.Quant.Grav. 18, 2339-2358  (2001),   \texttt{arXiv:hep-th/0007137}.
\bibitem{NC2}
D. Berenstein, R. G.  Leigh, \textit{Non-commutative Calabi–Yau manifolds,} Phys. Review Letters, 84(20) 4737–4740 (2000).

\bibitem{NC3}  A. Belhaj, E. H. Saidi,  \textit{On Non Commutative Calabi-Yau Hypersurfaces}, 
  Phys.Lett. B523, 191-198  (2001),    \texttt{arXiv:hep-th/0108143}. 

\bibitem{NC4}   A. Belhaj, J. J. Manjarin, P. Resco, \textit{On Non-Commutative Orbifolds of K3 Surfaces},  J.Math.Phys. 44, 2507-2520  (2003),    \texttt{arXiv:hep-th/0207160}.

\bibitem{NC5}   A. Belhaj, J.  Rasmussen, E.  H. Saidi, A.  Sebbar, \textit{Non-commutative ADE geometries as holomorphic wave equations},   Nucl.Phys. B727,  499-512  (2005),   \texttt{arXiv:hep-th/0504049}.





\bibitem{1}  P.~S.~Letelier, \textit{Clouds of strings in general relativity}, Phys.\ Rev.\ D 20, 1294 (1979).

\bibitem{2} A.~A.~Araújo Filho, J.~R.~Nascimento, A.~Yu.~Petrov, P.~J.~Porfírio, A.~Övgün, \textit{Effects of non-commutative geometry on black hole properties}, Phys.\ Dark Univ. 46, 101630 (2024), \texttt{arXiv:2406.12015}.

\bibitem{3}  A.~A.~Araújo Filho, J.~R.~Nascimento, A.~Yu.~Petrov, P.~J.~Porfírio, A.~Övgün, \textit{Properties of an axisymmetric Lorentzian non-commutative black hole}, Phys.\ Dark Univ.\ 47, 101796 (2025), \texttt{arXiv:2411.04674 [gr-qc]}. 



\bibitem{4} J. A. V. Campos, M. A. Anacleto, F. A. Brito, E. Passos, \textit{Quasinormal modes and
shadow of noncommutative black hole,} Sci. Rep. 12, 8516 (2022).
\bibitem{5} B. Hamil, B. C. Lütfüoğlu, \textit{Phantom RN-AdS black holes in noncommutative space,} Eur.
Phys. J. C 85, 313 (2025), \texttt{arXiv:2502.20514 [hep-th].}

\bibitem{N100} P. Nicolini, \textit{Noncommutative Black Holes, The Final Appeal To Quantum Gravity: A Review,} nt.J.Mod.Phys.A 24, 1229 (2009), \texttt{	arXiv:0807.1939 [hep-th].} 
\bibitem{9} F. Ahmed, A. R. P. Moreira, A. Bouzenada, \textit{Noncommutative Geometry Inspired
AdS Black Hole with a Cloud of Strings Surrounded by Quintessence-like fluid,}
\texttt{arXiv:2508.00740.}
\bibitem{10} G. Mascher, K. Destounis and K. D. Kokkotas, Charged black holes in de Sitter space:
superradiant amplification of charged scalar waves and resonant hyperradiation, Phys.
Rev. D105, 084052 (2022), \texttt{arXiv:2204.05335.}
\bibitem{11} P. Nicolini, A. Smailagic, E. Spallucci, \textit{Noncommutative geometry inspired Schwarzschild
black hole}, Phys. Lett. B 632, 547 (2006)

\bibitem{Q1} A. Tiwari, R. Kaur, J. K. Bhutto, T. Vayalpurayil, M. S. Habeeb, \textit{Criticality Quenching and Microstructure of Quintessence-AdS Black Holes}, \texttt{	arXiv:2605.12632 [hep-th]}.

\bibitem{f34} S. E. Baddis, A. Belhaj, H. Belmahi, M. Jemri, \textit{
Constraining Black Hole Shadows in Dunkl Spacetime using CUDA Numerical
Computations,} Journal of High Energy Astrophysics 51, 100541 (2025), \texttt{%
\ arXiv:2510.16460 [gr-qc]}.


\bibitem{s1} S. E. Baddis, A. Belhaj, and H. Belmahi,\textit{\ CUDA Assisted
Swampland and Black Hole Thermodynamics}, \texttt{arXiv:2508.12378 [hep-th]}.

\bibitem{s2} P. Berczik, R. Spurzem, L. Wang, S. Zhong, O. Veles, I.
Zinchenko, S. Huang, M. Tsai, G. Kennedy, S. Li, L. Naso, and C. Li, \textit{%
Up to 700k GPU cores, Kepler, and the Ex ascale future for simulations of
star clusters around black holes,} \texttt{arXiv:1312.1789 [astro-ph.IM].}

\bibitem{s3} A. G. M. Lewis, H. P. Pfeiffer, \textit{GPU-Accelerated
Simulations of Isolated Black Holes,} Class. Quant. Grav. 35, 095017 (2018), 
\texttt{arXiv:1804.09101 [gr-qc].}

\bibitem{s4} R. Ginjupalli and G. Khanna, \textit{High-Precision Numerical
Simulations of Rotating Black Holes Accelerated by CUDA,} \texttt{%
arXiv:1006.0663 [physics.comp-ph]. }

\bibitem{num1} W. H. Press et al., \textit{Numerical Recipes: The Art of Scientific Computing,} Cambridge University Press (2007).

\bibitem{num2} J. M. Stewart, \textit{Advanced General Relativity,} Cambridge University Press (1991).
\bibitem{num3} S. Chandrasekhar, \textit{The Mathematical Theory of Black Holes,} Oxford University Press (1983).

\bibitem{cuda1} J. Nickolls and I. Buck, \textit{CUDA: Scalable Parallel Programming Model for High-Performance Computing,} ACM Queue (2008).

\bibitem{cuda2} D. B. Kirk and W. Hwu, \textit{Programming Massively Parallel Processors: A Hands-on Approach,} Morgan Kaufmann (2016).



\bibitem{bhnum1} V. Perlick, \textit{Ray Optics, Fermat’s Principle, and Applications to General Relativity,} Springer (2000).

\bibitem{bhnum2} P. V. P. Cunha and C. A. R. Herdeiro, \textit{Stationary Black Holes and Light Rings,} Phys. Rev. Lett. 124, 181101 (2020), \texttt{	arXiv:2003.06445 [gr-qc]}.


\bibitem{HerdeiroRadu2014} C. A. R. Herdeiro, E. Radu, \textit{Kerr black holes with scalar hair,} Phys.Rev.Lett. 112, 221101 (2014), \texttt{	arXiv:1403.2757 [gr-qc].}

\bibitem{CunhaHerdeiro2018} P. V. P. Cunha, C. A. R. Herdeiro, \textit{Shadows and strong gravitational lensing: a brief review,} Gen.Rel.Grav. 50,  42 (2018), \texttt{		arXiv:1801.00860 [gr-qc].}

\bibitem{230} P. V. P. Cunha, C. A. R. Herdeiro, B. Kleihaus, J. Kunz, E. Radu, \textit{Shadows of Einstein-dilaton-Gauss-Bonnet black holes,} Phys. Lett. B 768, 373 (2017), \texttt{arXiv:1701.00079 [gr-qc].}

\bibitem{231} C. A. R. Herdeiro, E. Radu, \textit{Asymptotically flat black holes with scalar hair: a review,} Int. J. Mod. Phys. D 24, 1542014 (2015), \texttt{arXiv:1504.08209 [gr-qc].}

\bibitem{26} S. W. Wei, Y. C. Zou, Y. X. Liu, R. B. Mann, \textit{Curvature
radius and Kerr black hole shadow}, JCAP 08, 030 (2019), \texttt{%
arXiv:1904.07710.}

\bibitem{43} A. Elafrou, G. Thomas Collignon, \textit{Introduction to CUDA
Performance Optimization,} Nvidia.

\bibitem{44} Nvidia, \textit{CUDA C++ Programming Guide.}


\bibitem{27} Y. Décanini, A. Folacci, G. Esposito-Farèse,
\textit{Universality of high-energy absorption cross sections for black holes},
Phys.\ Rev.\ D 83, 044032 (2011),
\texttt{arXiv:1101.0781 [gr-qc]}.

\bibitem{28} S.-W. Wei, Y.-X. Liu, \textit{Relationship between high-energy
absorption cross section and strong gravitational lensing for a static and
spherically symmetric black hole}, Phys. Rev. D 84, 041501 (2011), \texttt{%
arXiv:1103.3822 [hep-th]}.

\bibitem{29} V. Perlick, \textit{Calculating black hole shadows: Review of
analytical studies,} Phys. Rep. 924, 1 (2022), \texttt{arXiv:2105.07101
[gr-qc]}.

\bibitem{E1} P. Kocherlakota et al. [Event Horizon Telescope],\textit{\
Constraints on black-hole charges with the 2017 EHT observations of M87*},
Phys. Rev. D 103, 10, 104047 (2021).

\bibitem{E2} L. Chakhchi, H. El Moumni and K. Masmar, \textit{Signatures of
the accelerating black holes with a cosmological constant from the Sgr A*
and M87* shadow prospects}, Phys. Dark Univ. 44, 101501 (2024).

\bibitem{E3} D. J. Gogoi and S. Ponglertsakul, \textit{Constraints on
quasinormal modes from black hole shadows in regular non-minimal Einstein
Yang Mills gravity}, Eur. Phys. J. C 84, 6652 (2024).
\bibitem{E34} N. Steinle, S. Safi-Harb, \textit{Machine learning classification of black holes in the
mass--spin diagram,} Phys. Rev. D112, 10, 103038 (2025), \texttt{arXiv:2508.14316.}
\bibitem{E35} J. W. Lee, Z. Y. Kim, \textit{Black hole/quantum machine learning correspondence,}
\text{arXiv:2506.09678.}
\bibitem{E36} W. Cui, X. Gao, M. Karkheiran, J. Wang, \textit{Machine Learning Free Quotients of CICYs,}
\texttt{arXiv:2508.19157 [hep-th].}
\bibitem{E37} L. F. S. Scabini, O. M. Bruno, \textit{Structure and Performance of Fully Connected Neural
Networks: Emerging Complex Network Properties,} \texttt{arXiv:2107.14062.}
\bibitem{ML1} J. Carifio, J. Halverson, D. Krioukov, B. D. Nelson, \textit{Machine Learning in the String Landscape}, JHEP 09, 157 (2017), \texttt{arXiv:1707.00655 [hep-th].}
\bibitem{ML2} Y.-H. He, \textit{Deep-Learning the Landscape}, Phys. Lett. B 774,  564 (2017), \texttt{arXiv:1706.02714 [hep-th].}

\end{thebibliography}
\end{document}